\documentclass[journal]{IEEEtran}
\usepackage{cite}
\usepackage{graphicx}
\usepackage[justification=centering]{caption}
\usepackage{autobreak}
\usepackage{setspace}

\ifCLASSINFOpdf
\else
\fi
\usepackage{amsmath}
\begin{document}
\title{Two New Kinds of Interference Alignment Schemes for Cellular $K$-user MIMO Downlink Networks}

\author{Jingfu~Li,~\IEEEmembership{}
        Wenjiang~Feng,~\IEEEmembership{}
        F. Richard Yu,~\IEEEmembership{Fellow,~IEEE}  
        and Weiheng~Jiang~\IEEEmembership{member,~IEEE}
\thanks{Jingfu Li, Wenjiang Feng and Weiheng Jiang are with the School of Microelectronics and Communication Engineering, Chongqing University, Chongqing 400044, China (e-mail:jingfuli@cqu.edu.cn; fengwj@cqu.edu.cn; whjiang@cqu.edu.cn ).}% <-this % stops a space
\thanks{Fei Richard Yu is with the Department of System and Computer
Engineering, Carleton University, Ottawa, ON K1S 5B6, Canada (e-mail: richard.yu@carleton.ca).}}% <-this % stops a space



\markboth{-}%
{-}




\maketitle

\begin{abstract}
It is known that interference alignment (IA) plays an important role in improving the degree of freedom (DoF) of multi-input and multi-output (MIMO) systems. However, most of the traditional IA schemes suffer from the high computational complexity and require the global and instantaneous channel state information (CSI), both of which make them difficult to be extended to cellular MIMO systems. To handle these issues, two new interference alignment schemes, i.e., the retrospective interference regeneration (RIR) scheme and the beamforming based distributed retrospective interference alignment (B-DRIA) scheme, are proposed for cellular $K$-user MIMO downlink networks. For the RIR scheme, it adopts interference elimination algorithm to erase redundant symbols in inter-cell interference (ICI) signals, and then uses interference regeneration algorithm to avoid secondary interference. The RIR scheme obtains greater DoF gain than the retrospective interference alignment (RIA) scheme, but incurs performance degradation when the transceiver antennas ratio (TAR) approaches 1. Therefore, the B-DRIA scheme is further proposed. For the B-DRIA scheme, the cellular beamforming matrix is introduced to eliminate the ICI, and meanwhile distributed retrospective interference alignment algorithm is adopted to align inter-user interference (IUI). The simulation results show that the B-DRIA scheme obtains larger DoF than the RIR scheme locally. Specifically, when TAR approaches 1, two schemes obtain the same DoF. While  TAR approaches 2, the DoF of the B-DRIA scheme is superior than the RIR scheme.
\end{abstract}


% \fi
\IEEEpeerreviewmaketitle

\section{Introduction}
\IEEEPARstart{T}{he} interference alignment (IA) technique is an elegant way towards improving the degree of freedom (DoF) of MIMO systems. With the help of IA technique, the $K$-user time-varying interference channel almost surely has ${K \mathord{\left/{\vphantom {K 2}} \right. \kern-\nulldelimiterspace} 2}$ DoF \cite{2008Interference}. However, the DoF gain caused by IA technique comes from the cost of perfect instantaneous channel state information at the transmitter (CSIT) acquisition \cite{2008Interference2,Shin2011On,Lee2013Uplink,2013Degrees}, which is a critical challenge, especially for distributed cellular networks \cite{Love2008An,2011Downlink,2011Interference}. That is, in distributed networks, the CSI feedback from receivers to transmitters experiences unavoidable delay, and meanwhile CSIT sharing among multiple transmitters will take up lots of feedback resources and bring burden to transmitters. To handle these issues, different kinds of approaches have been proposed, such as the blind interference alignment (BIA), the space-time interference alignment (STIA) and the retrospective interference alignment (RIA).

Concerning the broadcast channel (BC), a novel approach is to take IA technique without CSIT, which is the BIA  \cite{2012Blind}. Under certain heterogeneous block fading models, the BIA makes use of channel characteristics to keep channel gain constant in adjacent slots, and meanwhile alters the values of signals to reduce the dimension of interference signals. However, since the real channel fading model is somewhat between independent distributed fading model and block fading model, the scheme still cannot work in practical application. Therefore, for the $K$-user MISO channel systems with reconfigurable antennas, a new blind interference alignment is presented in \cite{2011Aiming,5962838}. Owing to the circulant and non-circulant structures of the scheme, the channels over different users keep correlated, which means that the inter-user interference cancellation and inter-subblock interference cancellation can be used to reduce the dimension of interference signals. Subsequently, the BIA gets extended to other systems, i.e., the cellular interference systems \cite{2013Blind} and heterogeneous systems \cite{Zhou2012On}. Nevertheless, the BIA still has the defect that the complexity of precoding matrix increases exponentially with the number of user antennas, which makes it hard to be used in MIMO systems.

To make IA technique practical and applicable in MIMO systems, the idea of making use of the moderate delayed CSIT is taken into account. Specifically, \cite{2013Space} firstly proposes STIA scheme for the $K$-user MISO BC. The core idea of the STIA is that, in block fading model, the transmitter combines IA technique with physical network coding (PNC) to eliminate interference in space-time domain. Subsequent studies achieve extensions of the STIA to different application scenarios \cite{Lee2014Distributed,2013CSI,2013Space2}. In particular, \cite{Lee2014Distributed} puts forward distributed STIA (DSTIA) for $2 \times 2$ user SISO $X$-channel (XC) and $3 \times 3$ user SISO interference channel (IC) which achieves ${4 \mathord{\left/{\vphantom {4 3}} \right. \kern-\nulldelimiterspace} 3}$ and ${6 \mathord{\left/{\vphantom {6 5}} \right. \kern-\nulldelimiterspace} 5}$ DoF respectively. \cite{2013CSI} introduces DSTIA into the $K$-user MISO IC and achieves $K-1$ DoF. For $K$-user MIMO BC and $K$-user MIMO IC, \cite{2013Space2} analyzes the trade-off strategy between achievable DoF and the range of CSI delay. However, the DoF gain of the STIA is restricted to the time span of the feedback delay, i.e., the time span of the feedback delay cannot exceed the time span of the coherent time. At the same time, owing to the inverse operation of the precoding matrix, the scheme brings burden to the transceivers.

To handle the issue of feedback delay, the delayed CSIT is studied. The RIA is firstly proposed \cite{Maddah2012Completely} for MISO networks and gets extended to $K$-user MIMO networks \cite{2014Linear}. The core idea of the RIA is adopting the repetition coding, i.e., at the transmitter, partial data is repeatedly transmitted $R$ times during the slots, and then the received signals are jointly decoded at the receivers' side at last. Subsequently, the scheme is extended to the IC scenarios, i.e., SISO IC \cite{2011Achieving}, MISO IC \cite{2014On}, MIMO IC \cite{Hao2016Achievable}, MIMO BC \cite{2015Retrospective,2015Retrospective2}, and meanwhile the achievable DoF gets further studied. However, with different channels, the upper bounds of the DoF are quite different. On the one hand, \cite{6179997} explores the upper bounds of the DoF for 2-user MIMO BC and 3-user MIMO BC, where the bounds are tight to the DoF obtained by RIA. On the other hand, by taking RIA scheme, the obtained DoF is relaxed to the upper bounds in the $K$-user MIMO IBC \cite{6205390} and $2 \times 2$-user SISO XC \cite{6341083} for the sake of the distributed transmitter loss. Therefore, the TDMA groups (TG) scheme and 3-user PSR scheme are introduced to maintain the DoF gain made by RIA, under which the system can be adaptive with different kinds of system configurations \cite{7588140}. After making tradeoff among three schemes, unfortunately, the obtained DoF is still not optimal in the cellular system, because it cannot properly handle the issue of inter-cell interference (ICI) \cite{6388347} by the middle way.

From the above illustration, we know that, when the number of user antennas is large, the BIA suffers from high computational complexity so that its application is not practical in MIMO systems. The STIA is adopted in MIMO systems and gets DoF gain, but its performance is restricted to the time span of the feedback delay. The RIA scheme solves the issue of feedback delay, but the effect of DoF gain gets affected in cellular networks. Therefore, an interference alignment scheme suitable for cellular networks is desired. In this paper, two new interference alignment schemes, i.e., the retrospective interference regeneration (RIR) scheme and the beamforming based distributed retrospective interference alignment (B-DRIA) scheme, are proposed for K-user cellular MIMO downlink systems. The main contribution of this paper is three-fold as follows:
\begin{itemize}
\vspace{-0.3em}
\item The paper proposes the RIR scheme which adopts interference elimination algorithm to erase redundant symbols in inter-cell interference (ICI) signals, and then uses interference regeneration algorithm to avoid secondary ICI interference. The scheme achieves higher DoF gain than the RIA scheme, i.e., 
${{LM} \mathord{\left/
 {\vphantom {{LM} {(L + {1 \mathord{\left/
 {\vphantom {1 {\left\lfloor {{1 \mathord{\left/
 {\vphantom {1 {(\rho  - 1)}}} \right.
 \kern-\nulldelimiterspace} {(\rho  - 1)}}} \right\rfloor }}} \right.
 \kern-\nulldelimiterspace} {\left\lfloor {{1 \mathord{\left/
 {\vphantom {1 {(\rho  - 1)}}} \right.
 \kern-\nulldelimiterspace} {(\rho  - 1)}}} \right\rfloor }})}}} \right.
 \kern-\nulldelimiterspace} {(L + {1 \mathord{\left/
 {\vphantom {1 {\left\lfloor {{1 \mathord{\left/
 {\vphantom {1 {(\rho  - 1)}}} \right.
 \kern-\nulldelimiterspace} {(\rho  - 1)}}} \right\rfloor }}} \right.
 \kern-\nulldelimiterspace} {\left\lfloor {{1 \mathord{\left/
 {\vphantom {1 {(\rho  - 1)}}} \right.
 \kern-\nulldelimiterspace} {(\rho  - 1)}}} \right\rfloor }})}}$
, but incurs performance degradation when transceiver antennas ratio approaches 1.

\item To avoid the performance degradation of the RIR scheme under certain circumstances, the paper further proposes the B-DRIA scheme which adopts the cellular beamforming matrix to eliminate the ICI, and meanwhile utilizes distributed retrospective interference alignment algorithm to align inter-user interference (IUI). The scheme maintains the same performance improvement for the DoF, i.e.,
${{L\left\lceil {{M \mathord{\left/
 {\vphantom {M K}} \right.
 \kern-\nulldelimiterspace} K}} \right\rceil K(\bar \varphi { + }1)} \mathord{\left/
 {\vphantom {{L\left\lceil {{M \mathord{\left/
 {\vphantom {M K}} \right.
 \kern-\nulldelimiterspace} K}} \right\rceil K(\bar \varphi { + }1)} {(L{ + }\bar \varphi { + }}1)}} \right.
 \kern-\nulldelimiterspace} {(L{ + }\bar \varphi { + }1)}}$, and in the meantime avoids performance degradation of the RIR scheme.

\item To further analyze the performance of the two proposed schemes effected by transceiver antennas ratio, we perform numerical evaluations via simulations. The simulation results show that, when the transceiver antennas ratio approaches 2, both RIR scheme and B-DRIA scheme obtain greater DoF gain than the RIA scheme, while as the transceiver antennas ratio approaches 1, B-DRIA scheme achieves greater DoF gain than RIR scheme.
\vspace{-0.3em}
\end{itemize}

The rest of this paper is organized as follows. The cellular distributed multi-user MIMO system is presented in Section II, which includes the description of the network in time domain and angular domain, and meanwhile the CSIT feedback model and the performance criteria are introduced. Then, two novel interference alignment schemes, i.e., the RIR scheme and the B-DRIA scheme, are proposed and the corresponding typical applications are given in Section III and Section IV, respectively. The performances of the proposed schemes are evaluated in Section V and we conclude at last.

\section{System Model}
Consider cellular distributed $K$-user MIMO downlink network \cite{7464855} as illustrated in Fig. 1. Define the cells set as ${\cal L}{ = }\{ 1,2, \cdots ,L\}$, and for each cell $i,\forall i \in {\cal L}$, the base station is configured with $M$ antennas to provide service to the users in set 
${{\textbf{K}}_i}{ = }\left\{ {1, \cdots ,K} \right\}$. In addition, each served user $k,\forall k \in {{\textbf{K}}_i}$ in the cell $i$ is equipped with $N$ antennas where $M>N$, which means that no user can decode its message without additional information.
\begin{figure}[!t]
\centering
\includegraphics[width=2.5in]{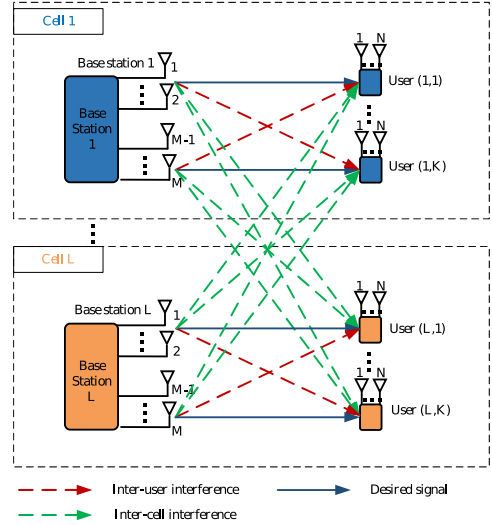}
\caption{Cellular distributed multi-user MIMO network.}
\label{Fig1}
\end{figure}

In the following, for the simplicity, we use $i,\forall i \in {\cal L}$ and $j,\forall j \in {\cal L}$ to denote the cell which the transmitter and the receiver belong to, respectively. Then, we further classify the cells into two categories: if $j = i$, the cell $i$ is called target cell, otherwise denotes  interference cell. Assume that the network takes ${\textbf{T}}{ = }\{ 1,2, \cdots ,T\}$ slots to transmit signals and for each slot $t,\forall t \in {\textbf{T}}$, the base station $i$ sends messages ${{\bf{S}}_i}[t] = \left[ {{{\bf{S}}_{i,1}}[t] \cdots {{\bf{S}}_{i,k}}[t]} \right] $ to its served users. Due to the broadcast nature of the wireless communication, on the one hand, the received signals of the user $k,\forall k \in{\textbf{K}}$ are composed of two parts, i.e., the disred signals ${{\bf{S}}_i}[t]{ = }\left[ {{{\bf{S}}_{i,1}}[t] \cdots {{\bf{S}}_{i,k}}[t]} \right]$ from the cell $i$ and undesired signals  ${{\bf{S}}_{ - i}}[t]{ = }\left[ {{{\bf{S}}_{ - i,1}}[t] \cdots {{\bf{S}}_{ - i,k}}[t]} \right]$ from the interference cells $ - i = {\cal L}\backslash i$. On the other hand, the desired signals is made up of desired symbols ${{\bf{S}}_{i,k}}[t]{ = }\left[ {{s_{k,1}}[t] \cdots {s_{k,n}}[t]} \right]$ where $0\!\! <\!n \!\!\le\! N$ and undesired symbols ${{\bf{S}}_{i, - k}}[t]{ = }\left[ {{s_{ - k,1}}[t] \cdots {s_{ - k,n}}[t]} \right]$ where $ - k ={\textbf{K}} \backslash k$.

Herein, for the cellular distributed K-user MIMO downlink network, since all users share the same channel resources, i.e., the IUI and the ICI coexist in each slot. In order to implement our schemes to align IUI and ICI, the CSI should be further differentiated. Specifically, the component of the CSI should be expanded into two parts, the channel gain information and the angel information. Thus, the system model should be described in time domain \cite{5358700} and angular domain \cite{1143830}, as that presented in the section II.A and II.B, respectively. Meanwhile, based on the relationship between the coherent time and the time delay, the types of the CSI should be classified into three categories, the instantaneous CSI, the moderate delayed CSI and the delayed CSI. In specific, the CSIT feedback model is presented in the section II.C \cite{6926832}.

\subsection{Time Domain Model}
To acquire the channel gain information of CSI, the system model mentioned above should be described as the time domain model \cite{5358700}. Taking the transmission of one slot $t,\forall t \in {\textbf{T}} $ as example, the base station $i$ transmits messages ${\boldsymbol{{S}}_{i,k}}[t]$ to the user $k$ in the cell $i$. Then the received signal of the user $k$ in cell $j,\forall j \in {\cal L}$ can be written by 
\begin{equation}
\label{eq1}
{\boldsymbol{{y}}_{j,k}}[t]{ = }\sum\limits_{i = 1}^L {\sum\limits_{k = 1}^K {{\bf{H}}_i^{j,k}[t]{{\boldsymbol{{S}}}_{i,k}}[t]} } { + }\overline {\boldsymbol{{N}}}.
\end{equation}
where ${\bf{H}}_i^{j,k}[t] \in{\textbf{C}}{^{M \times N}}$ denotes the channel matrix between the base station $i$ and the user $k$ of the cell $j$, and it is subject to independent and identically distributed $\left( {i.i.d.} \right)$ according to ${\cal C}{\cal N}(0,1)$. $\overline {\boldsymbol{N}}  \in{\textbf{C}}{^{N \times 1}}$ denotes the additive white Gaussian noise (AWGN) term at the user $k$ of the cell $j$.

If the transmitted messages ${{\boldsymbol{S}}_{i,k}}[t]$ is known to the user $k$ in the cell $i$, the channel matrix ${\bf{H}}_i^{j,k}[t]$ can be gotten in the form of the time domain, which denotes the channel gain information of CSI.

\subsection{Angular Domain Model}
Similarly, to acquire the angel information of CSI, the system model of the network should be described in angular domain \cite{1143830}. The application scenario is the same as it is assumed in the section II.A. The received signals are given by
\begin{equation}
\label{eq2}
{{\boldsymbol{y}}_{j,k}}[t]{ = }\sum\limits_{i = 1}^L {\sum\limits_{k = 1}^K {{\bf{A}}_i^{j,k}[{\theta _i}]{{\boldsymbol{S}}_{i,k}}[t]} } { + }\overline {\boldsymbol{N}}.
\end{equation}
where ${\bf{A}}_i^{j,k}[{\theta _i}] \in{\textbf{C}}{^{N \times M}}$  is the direction of arrival (DoA) matrix between the base station $i$ and the user $k$ in the cell $j$ and meanwhile, $\overline {\boldsymbol{N}}\in{\textbf{C}}{^{N \times 1}}$ denotes AWGN. By analyzing the ingredient of ${\bf{A}}_i^{j,k}[{\theta _i}]$, it is composed of $M$ direction vectors,
\begin{equation}
\label{eq3}
{\bf{A}}_i^{j,k}[{\theta _i}]{ = }\left[ {\begin{array}{*{20}{c}}
{{{\boldsymbol{a}}}_{i,1}^{j,k}[{\theta _i}]}& \cdots &{{{\boldsymbol{a}}}_{i,M}^{j,k}[{\theta _i}]}
\end{array}} \right],
\end{equation}
where the direction vector ${{\boldsymbol{a}}}_{i,m}^{j,k}[{\theta _i}]$ can be regarded as the function of the DoA angular ${\theta _i}$. As for each ${{\boldsymbol{a}}}_{i,m}^{j,k}[{\theta _i}]$, it represents the direction from the antenna $m \in [1,M]$ of the cell i to the user $k$ in the cell $j$ and is written as
\begin{equation}
\label{eq4}
{\boldsymbol{a}}_{i,m}^{j,k}[{\theta _i}] = {\left[ {1,{e^{ - j{\mu _{m}}}}, \cdots ,{e^{ - j(N - 1){\mu _m}}}} \right]^T}.
\end{equation}
Note that ${\mu _m} = ({{2\pi d} \mathord{\left/{\vphantom {{2\pi d} \lambda }} \right.\kern-\nulldelimiterspace} \lambda })\sin {\theta _{i}}$, where $d$ is the array spacing and $\lambda $ is the carrier wavelength. The DoA angular ${\theta _i}$ denotes the angel information of CSI.

\subsection{CSIT Feedback Model}
Due to the unavoidable feedback delay from the receiver to the transmitter, after making expansion of the component of CSI, the obtained CSIT is a delayed version. It means that, at the transmitter side, the obtained CSIT cannot be used to instruct the design of precoding matrix at the current slot. Therefore, the types of the CSI should be further classified by CSIT feedback model \cite{6926832} which is summarized in Fig. 2. 
\begin{figure}[!t]
\centering
\includegraphics[width=2.5in]{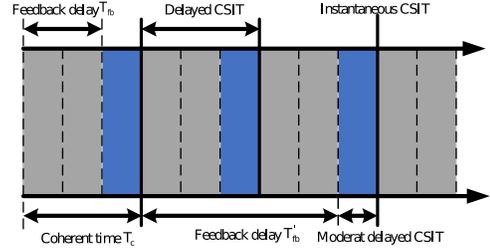}
\caption{CSIT feedback model.}
\label{Fig2}
\end{figure}

We assume that the receiver can perfectly estimate the CSI and then send it back to the corresponding transmitter through an error-free but delayed feedback link. Therefore, the transmitter can continuously track the variation of the channel matrix over different slots. To distinguish different versions of CSIT, we define the channel feedback delay and the period of coherence time as ${T_{fb}}$ and ${T_c}$, respectively. Then, the definition of the ratio $\lambda$ is given by
\begin{equation}
\label{eq5}
\lambda { = }{{{T_{fb}}} \mathord{\left/
 {\vphantom {{{T_{fb}}} {{T_c}}}} \right.
 \kern-\nulldelimiterspace} {{T_c}}},
\end{equation}
which is naturally divided into three types and each type represents one kind of CSI:
\begin{itemize}
\vspace{-0.3em}
\item $\lambda  = 0$, the obtained CSI is an instantaneous CSI, that is the transmitters acquire the CSI feedback from the receivers immediately.

\item $0 < \lambda  < 1$, this situation should be subdivided into two subcases, i.e., the moment of getting CSI belongs to $\left[ {{T_{fb}} ,{T_c}- {T_{fb}}} \right]$ and $\left[ {{T_c} - {T_{fb}},{T_c}} \right]$. In the former interval, the obtained CSI is equivalent to the instantaneous CSI, while in the later interval, the obtained CSI represents the moderate delayed CSI, which means that, after feedback delay, the transmitters can also make use of the CSI in the rest time of the current slot.

\item $\lambda  \ge 1$, the obtained CSI is a delayed version, that is the CSI is absolutely outdated. Thus, the channel gain for the current slot is unknown and the CSI is no longer limited by the feedback delay.
\vspace{-0.3em}
\end{itemize}
The type of CSI adopted in Section III and Section IV should be a comprehensive combination of the latter two types, i.e., the moderate delayed CSI and the delayed CSI.

\subsection{Sum Degrees of Freedom}
For the considered system, a measurement criterion is desired to compare the performance of the proposed schemes with existing IA schemes. Herein, the sum degrees of freedom \cite{7442513} is extended to the cellular distributed $K$-user MIMO downlink network. Specifically, in each slot $t,\forall t \!\in\!{\textbf{T}}$, the base stations simultaneously transmit independent messages. Taking the base station $i,\forall i \in {\cal L}$ for example, the messages ${{\bf{S}}_{i,1}}[t], \cdots ,{{\bf{S}}_{i,k}}[t]$ are sent with the rates of ${R_{i,1}}, \cdots ,{R_{i,k}}$  bits/s/Hz, respectively. For the user $k,\forall k \in {\textbf{K}}$, its achievable rate is characterized by 
\begin{equation}
\label{eq6}
{R_{i,k}}{ = }{d_{i,k}}\log (1{ + }SNR) + o\left( {\log (SNR)} \right),
\end{equation}
where ${d_{i,k}}$ denotes the degree of freedom belonging to the user $k$ in the cell $i$ and can be expressed as
\begin{equation}
\label{eq7}
{d_{i,k}} = \mathop {\lim }\limits_{SNR \to \infty } {{{R_{i,k}}} \mathord{\left/
 {\vphantom {{{R_{i,k}}} {{{\log }_2}(SNR)}}} \right.
 \kern-\nulldelimiterspace} {{{\log }_2}(SNR)}}.
\end{equation}
The formula (7) measures the channel capacity of a single cell per user, i.e., the number of data streams that can be transmitted reliably by one user within one slot. Then, we extend it to the case of cellar $K$-user, given by
\begin{equation}
\label{eq8}
DoF = \mathop {\lim }\limits_{SNR \to \infty } \mathop {\max }\limits_{\boldsymbol{R}} \sum\limits_{i = 1}^L {\sum\limits_{k = 1}^K {{{{R_{i,k}}} \mathord{\left/
 {\vphantom {{{R_{i,k}}} {{{\log }_2}(SNR)}}} \right.
 \kern-\nulldelimiterspace} {{{\log }_2}(SNR)}}} },
\end{equation}
where ${\boldsymbol{R}}{ = }\left[ {{R_{i,1}}, \cdots ,{R_{i,k}}} \right] \in {{\textbf{R}}^K}$ denotes the vector of achievable rates for the users in the cell $i$. Thus, the DoF measures the channel capacity of the whole network.

\section{Retrospective Interference Regeneration Scheme}
Although the RIA scheme achieves greater DoF than TDMA scheme 
\cite{Maddah2012Completely,2014Linear,2011Achieving,2014On,Hao2016Achievable,2015Retrospective,2015Retrospective2,6179997,6205390,6341083,7588140,6388347}, when the numbers of cells and users increase, some space resources are wasted and the DoF gain will be decreased. To handle this issue, we propose the retrospective interference regeneration (RIR) scheme. The core idea of the scheme is making use of the interference signals from interference cells to align ICI, i.e., when the receivers feed interference signals back, the base stations eliminate redundant symbols in the interference signals to extract desired symbols and meanwhile, provide desired symbols to the target cell. In the following, we firstly illustrate the details of the proposed scheme which includes three phases, i.e.,  signal transmission, interference elimination and interference regeneration, and  interference retransmission. Then, the DoF of the RIR scheme is analyzed. To make it understandable, we also present an example of the scheme at last. 

\subsection{Signal Transmission}
The ﬁrst phase spans ${\cal L}$ groups of $\varphi  = \left\lfloor {{N \mathord{\left/ {\vphantom {N {\left( {M - N} \right)}}} \right. \kern-\nulldelimiterspace} {\left( {M - N} \right)}}} \right\rfloor $ slots where $\left\lfloor {*} \right\rfloor $ denotes the round down operation. In each slot $t \in \left\{ {1, \cdots ,\varphi } \right\}$ of period $i \in {\cal L}$, the base station $i$ transmits signals while the other base stations keep silent. The transmitted signal is characterized by
\begin{equation}
\label{eq9}
{\boldsymbol{S}}_{i}^{}[(i - 1)\varphi  + t] = {\left[ {{s_{i,1}}[(i - 1)\varphi  + t], \cdots ,{s_{i,M}}[(i - 1)\varphi  + t]} \right]^T}.
\end{equation}
Then the signals received by user $k$ of the cell $j$ are 
\begin{equation}
\label{eq10}
{\boldsymbol{y}}_{j,k}^{}[(i - 1)\varphi  + t] = {\bf{H}}_i^{j,k}[(i - 1)\varphi  + t]{{\boldsymbol{S}}_i}[(i - 1)\varphi  + t],j \in {\cal L}.
\end{equation}
As mentioned earlier, if $i \ne j$, the cell $i$ is the interference cell and the corresponding received signals are ICI. If $i = j$ , the cell $i$ is the target cell but, for the user $k$, only part of ${{\boldsymbol{S}}_i}[(i - 1)\varphi  + t]$ are desired symbols, i.e., 
\begin{small}
$\left\{ {{s_{i,(k - 1)\left\lceil {{M \mathord{\left/ {\vphantom {M K}} \right. \kern-\nulldelimiterspace} K}} \right\rceil  + 1}}\left[ {(i - 1)\varphi  + t} \right], \cdots ,{s_{i,k\left\lceil {{M \mathord{\left/ {\vphantom {M K}} \right. \kern-\nulldelimiterspace} K}} \right\rceil }}\left[ {(i - 1)\varphi  + t} \right]} \right\}$
\end{small}
 and the left are the interference symbols. According to whether this user belong to the target cell, the user feedback different information to the transmitter, i.e., if $i = j$, the users feed the channel gain information back to the target cell’s base station by delayed feedback links. While if $i \ne j$, the users feed the ICI back to the interference cell’s base station, i.e., ${\boldsymbol{y}}_{j,k}^{}[(i - 1)\varphi  + t]$. This process is illustrated by Fig. 3.
\begin{figure}[!t]
\centering
\includegraphics[width=2.5in]{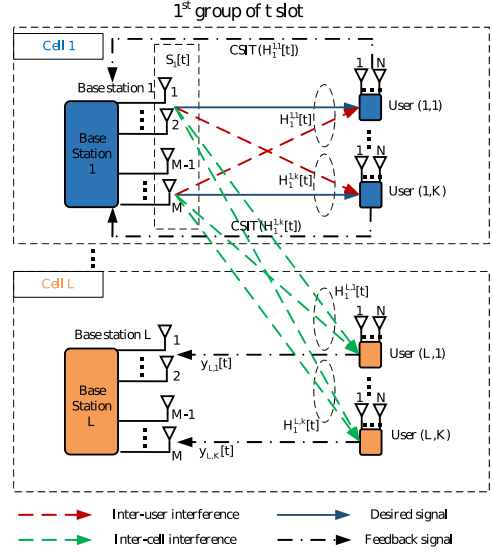}
\caption{Signal transmission in 1st phase.}
\label{Fig3}
\end{figure}

As mentioned earlier, since $M > N$, no user can decode the messages with only one-slot transmitted signals. To handle this issue, the additional interference alignment process is developed to the system, i.e., interference elimination and interference regeneration.

\subsection{Interference Elimination and Interference Regeneration}
By analyzing the components of ICI, we find that, for the user $k$, part of the information in ICI can be utilized, while the remainder is still the interference. Therefore, to distinguish these two parts, the ICI can be written as 
\begin{align}
\label{eq11}
{{\textbf{L}}_j}[(i\! -\! 1)\varphi  \!+ \!t]{ = }&{{\textbf{L}}_j}\left( {{s_{i,1}}[(i\! -\! 1)\varphi \! + \!t],\! \cdots\! ,{s_{i,M}}[(i\! - \!1)\varphi \! + \!t]} \right)\nonumber\\
{ = }&{\bf{H}}_i^{j,k}[(i - 1)\varphi  + t]{{\boldsymbol{S}}_i}[(i - 1)\varphi  + t],\nonumber\\
          &{\cal L},j \ne i,k \in {\textbf{K}}\nonumber\\
{ = }&{\boldsymbol{L}}_j^{i,k}[(i - 1)\varphi  + t]{ + }{\boldsymbol{I}}_j^{i,k}[(i - 1)\varphi  + t],
\end{align}
where ${\boldsymbol{L}}_j^{i,k}[(i - 1)\varphi  + t]$ represents the useful part and ${\boldsymbol{I}}_j^{i,k}[(i - 1)\varphi  + t]$ represents the rest, i.e., the interference. Then, we can define these two parts as partial desired signals and partial interference signals as follows, respectively,
\begin{align}
\label{eq12}
{\boldsymbol{L}}_j^{i,k}[(i - 1)\varphi  + t]
{ = }& {\boldsymbol{L}}_j^{i,k}({s_{i,(k - 1)\left\lceil {{M \mathord{\left/{\vphantom {M K}} \right.
            \kern-\nulldelimiterspace} K}} \right\rceil  + 1}}[(i - 1)\varphi  + t], \nonumber\\
          &\cdots ,{s_{i,k\left\lceil {{M \mathord{\left/{\vphantom {M K}} \right.\kern-\nulldelimiterspace} K}} \right\rceil }}[(i - 1)\varphi  + t])\nonumber\\
{=}  &{\bf{H}}_i^{j,k}[(i - 1)\varphi  + t]{{\boldsymbol{S}}_i}[(i - 1)\varphi  + t], \nonumber\\
          &j \in {\cal L},j \ne i,
\end{align}
\begin{align}
\label{eq13}
{\boldsymbol{I}}_j^{i,k}[(i - 1)\varphi  + t]
{ = }&{\boldsymbol{I}}_j^{i,k}({s_{i,( - k - 1)\left\lceil {{M \mathord{\left/{\vphantom {M K}} \right.
 \kern-\nulldelimiterspace} K}} \right\rceil  + 1}}[(i - 1)\varphi  + t],\nonumber\\
          &\cdots ,{s_{i, - k\left\lceil {{M \mathord{\left/{\vphantom {M K}} \right.\kern-\nulldelimiterspace} K}} \right\rceil }}[(i - 1)\varphi  + t])\nonumber\\
{ = }&{\bf{H}}_i^{j,k}[(i - 1)\varphi  + t]{{\boldsymbol{S}}_i}[(i - 1)\varphi  + t],\nonumber\\
          &j \in {\cal L},j \ne i, - k ={\textbf{K}} \backslash k,
\end{align}
For the interference cell, the interference space of ${\boldsymbol{L}}_j^{i,k}[(i - 1)\varphi  + t]$ is a subspace of ${{\textbf{L}}_j}[(i - 1)\varphi  + t]$. Hence, according to the feedback interference signals, the interference cell’s base station can design a precoding matrix ${{\bf{U}}_j}[(i - 1)\varphi  + t]$ to achieve partial interference elimination, i.e.,
\begin{equation}
\label{eq14}
{{\bf{U}}_j}[(i - 1)\varphi  + t]{{\textbf{L}}_j}[(i - 1)\varphi  + t] = {\boldsymbol{L}}_j^{i,k}[(i - 1)\varphi  + t].
\end{equation}
For (14), according to the extension theorem of linear subspace \cite{1998Matrix}, the precoding matrix ${{\bf{U}}_j}[t]$ must exist, and the solution of the matrix can be obtained by cyclic-zero-padding precoding matrix \cite{7820128}.

After the interference elimination, the partial desired signal ${\boldsymbol{L}}_j^{i,k}[(i - 1)\varphi  + t]$ can be transmitted in the interference retransmission phase. Unfortunately, the interference retransmission will cause additional interference to other cells except the target cell, as illustrated in Fig. 4.
\begin{figure}[!t]
\centering
\includegraphics[width=2.5in]{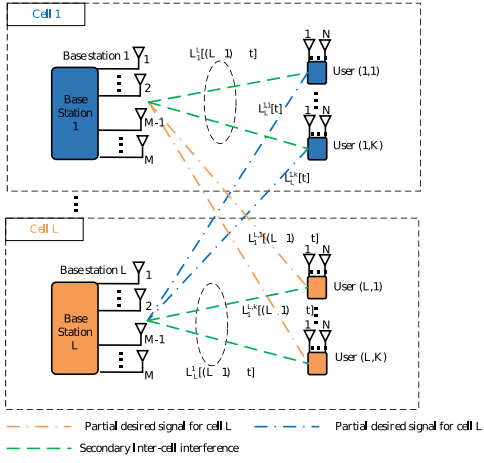}
\caption{Additional ICI in 2nd phase.}
\label{Fig4}
\end{figure}
 To handle this issue, the base station should design interference regeneration matrices to satisfy the following condition
\begin{equation}
\label{eq15}
\begin{array}{l}
{{\bf{V}}_p}[(i - 1)\varphi  + t]{\boldsymbol{L}}_p^{i,k}[(i - 1)\varphi  + t] \\
{ = }{{\bf{V}}_q}[(i - 1)\varphi  + t]{\boldsymbol{L}}_q^{i,k}[(i - 1)\varphi  + t],\forall p,q \in {\cal L}\backslash i,
\end{array}
\end{equation}
where each interference regeneration matrix ${{\bf{V}}_p}[(i - 1)\varphi  + t]$ is an $\left\lceil {{M \mathord{\left/
 {\vphantom {M K}} \right.
 \kern-\nulldelimiterspace} K}} \right\rceil  \times \left\lceil {{M \mathord{\left/
 {\vphantom {M K}} \right.
 \kern-\nulldelimiterspace} K}} \right\rceil $
 matrix that can align the interference signals into the same space. Since the vectors ${\boldsymbol{L}}_p^{i,k}[(i - 1)\varphi  + t]$ and ${\boldsymbol{L}}_q^{i,k}[(i - 1)\varphi  + t]$ are irrelevant with same dimension $\left\lceil {{M \mathord{\left/{\vphantom {M K}} \right.\kern-\nulldelimiterspace} K}} \right\rceil $, the precoding matrices ${{\bf{V}}_p}[(i - 1)\varphi  + t]$  and ${{\bf{V}}_q}[(i - 1)\varphi  + t]$ can be regarded as row transformation matrices. Thus, from the property of equivalence matrix \cite{1998Matrix},  the interference regeneration matrices are always exist, and cyclic-zero-padding precoding \cite{7820128} can be used to obtain the particular solution. The details are summarized in Fig. 5.
\begin{figure}[!t]
\centering
\includegraphics[width=2.5in]{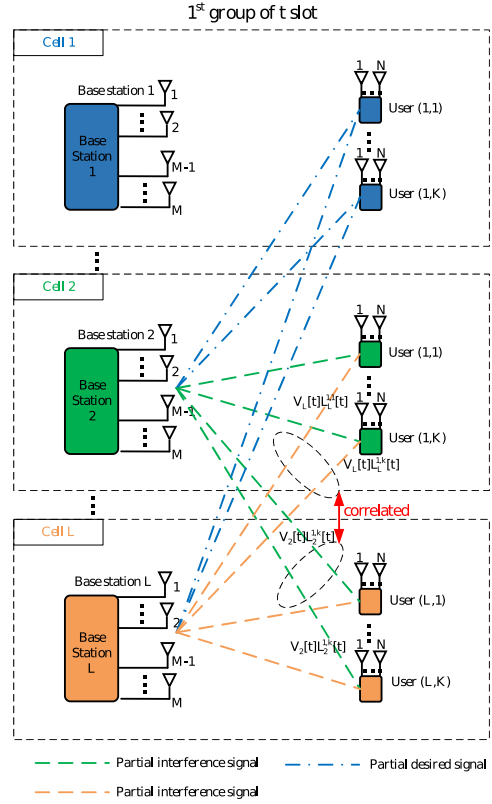}
\caption{Interference regeneration in 2nd phase.}
\label{Fig5}
\end{figure}

\subsection{Interference Retransmission}
After interference elimination and interference regeneration, base stations are capable to simultaneously transmit partial desired signals as 
\begin{align}
\label{eq16}
{{\boldsymbol{X}}_j}[L\varphi  + 1] &=\! \sum\nolimits_{i \in {\cal L}\backslash j}\! {{\boldsymbol{X}}_j^i[L\varphi  + 1]}\nonumber \\
                            &= \! \sum\nolimits_{i \in {\cal L}\backslash j} \! \! {\sum\limits_{t = \left( {i - 1} \right)\varphi  + 1}^{i\varphi } \! {{{\bf{V}}_j}[t]{{\bf{U}}_j}[t]({\bf{H}}_i^{j,k}[t]{{\boldsymbol{S}}_i}[t])} }\nonumber \\
                            &= \sum\nolimits_{i \in {\cal L}\backslash j} {\sum\limits_{t = \left( {i - 1} \right)\varphi  + 1}^{i\varphi } {{{\bf{V}}_j}[t]{\boldsymbol{L}}_j^{i,k}[t]} }.
\end{align}
With the preprocessing of ${{\bf{V}}_j}[t]$ and ${{\bf{U}}_j}[t]$, for the whole cells, the additional signals coming from the interference retransmission phase are convert to the partial desired signals, which means that the retransmitted process will not bring about ICI and IUI, as illustrated in Fig. 6. Herein, for the user $k$ in the target cell $i$, the received signals are written as,
\begin{equation}
\label{eq17}
{{\bf{y}}_{i,k}}[L\varphi  + 1] = \sum\nolimits_{j \in {\cal L}} {{\bf{H}}_j^{i,k}[L\varphi  + 1]{{\bf{X}}_j}[L\varphi  + 1]},
\end{equation}
\begin{figure}[!t]
\centering
\includegraphics[width=2.5in]{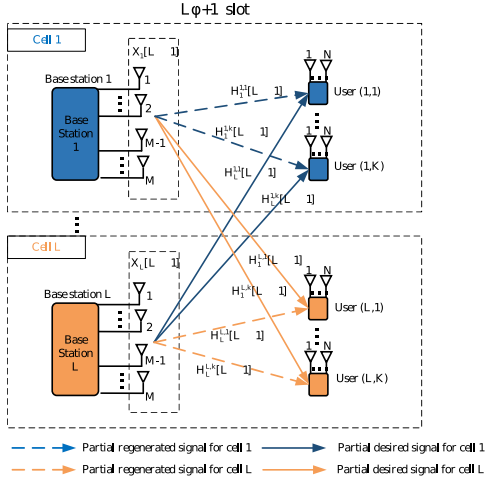}
\caption{Interference retransmission in 3rd phase.}
\label{Fig6}
\end{figure}
\!and the whole process can be summarized as the equation (18),
\newcounter{mytempeqncnt}
\begin{figure*}[!t]
\normalsize
\setcounter{mytempeqncnt}{\value{equation}}
\setcounter{equation}{17}
\begin {small}
\begin{equation}
\setlength{\arraycolsep}{0.5pt}
\label{eq18}
\underbrace {\left[ {\begin{array}{*{20}{c}}
{{{\bf{y}}_{i,k}}[(i - 1)\varphi  + 1]}\\
{{{\bf{y}}_{i,k}}[(i - 1)\varphi  + 2]}\\
 \vdots \\
{{{\bf{y}}_{i,k}}[i\varphi ]}\\
{{{\bf{y}}_{i,k}}[L\varphi  + 1]}
\end{array}} \right]}_{\bf{Y}}\! { = }\! \underbrace {\left[ {\begin{array}{*{20}{c}}
{{\bf{H}}_i^{i,k}[(i - 1)\varphi  + 1]}&0& \cdots &0\\
0& \ddots & \ddots & \vdots \\
 \vdots & \ddots & \ddots &0\\
0& \cdots &0&{{\bf{H}}_i^{i,k}[i\varphi ]}\\
{\sum\limits_{j \in {\cal L}} {{\bf{H}}_j^{i,k}[L\varphi  + 1]} {{\boldsymbol{\beta }}_{(i - 1)\varphi  + 1}}}&{\sum\limits_{j \in {\cal L}} {{\bf{H}}_j^{i,k}[L\varphi  + 1]{{\boldsymbol{\beta }}_{(i - 1)\varphi  + 2}}} }& \cdots &{\sum\limits_{j \in {\cal L}} {{\bf{H}}_j^{i,k}[L\varphi  + 1]{{\boldsymbol{\beta }}_{i\varphi }}} }
\end{array}} \right]}_{\bf{H}}\underbrace {\left[ {\begin{array}{*{20}{c}}
{{{\bf{S}}_i}[(i - 1)\varphi  + 1]}\\
 \vdots \\
 \vdots \\
 \vdots \\
{{{\bf{S}}_i}[i\varphi ]}
\end{array}} \right]}_{\bf{S}}
\end{equation}
\end {small}
\hrulefill
\vspace*{4pt}
\end{figure*}
 where ${{\boldsymbol{\beta}}_t}$ is the combination precoding of the interference cells. The precoding matrix is composed of interference elimination matrix, interference regeneration matrix and channel matrix in slot $t$, denoted by
\begin{equation}
\label{eq19}
{{\boldsymbol{\beta} }_t} = \sum\nolimits_{j \in {\cal L}{\backslash }i} {{{\bf{V}}_j}[t]{{\bf{U}}_j}[t]{\bf{H}}_i^{j,k}[t]}.
\end{equation}

For our proposed scheme, ${{\bf{V}}_j}[t]$ , ${{\bf{U}}_j}[t]$ and ${\bf{H}}_j^{i,k}[L\varphi  + 1]$ are non-singular matrix which means that they don't change the rank of ${\bf{H}}_i^{j,k}[t]$, and in the meantime, for any two subchannel matrices, they either  come from different time slots, e.g., ${\bf{H}}_i^{i,k}[(i - 1)\varphi  + 1]$ and ${\bf{H}}_i^{i,k}[i\varphi ]$, or come from different spaces, e.g., ${\bf{H}}_i^{i,k}[(i - 1)\varphi  + 1]$ and ${\bf{H}}_j^{i,k}[(i - 1)\varphi  + 1], j \ne i$, are independent. The former is for the sake of the delayed CSIT and the latter is on account of the irrelevance of the channels among different cells. Herein, the elements of $\bf{H}$ are independent so that the matrix itself is a $\left\lfloor {{N \mathord{\left/{\vphantom {N {(M - N}}} \right. \kern-\nulldelimiterspace} {(M - N}})} \right\rfloor M$ dimensional non-singular matrix. Therefore, at the user side, the Least Square (LS) method \cite{2004Discussion} can be used to decode the messages of the whole slots. 

\subsection{Degrees of Freedom}
In this subsection, the DoF of the proposed RIR scheme is analyzed. In specific, since the base stations spend $L\varphi { + }1$ incoherent slots to transmit $L\left\lceil {{M \mathord{\left/ {\vphantom {M K}} \right. \kern-\nulldelimiterspace} K}} \right\rceil K\varphi $ independent symbols in all. We assume that the transmission is reliable, i.e., the value of signal-to-noise ratio approaches infinity. Therefore, its achievable DoF is 
\begin{equation}
\label{eq20}
DoF{ = }{{(L\left\lceil {{M \mathord{\left/
 {\vphantom {M K}} \right.
 \kern-\nulldelimiterspace} K}} \right\rceil K\varphi )} \mathord{\left/
 {\vphantom {{(L\left\lceil {{M \mathord{\left/
 {\vphantom {M K}} \right.
 \kern-\nulldelimiterspace} K}} \right\rceil K\varphi )} {(L\varphi { + }1)}}} \right.
 \kern-\nulldelimiterspace} {(L\varphi { + }1)}}.
\end{equation}
To facilitate the analysis, we assume that $M$ can be divided by $K$. Then the DoF can be simplified as
\begin{equation}
\label{eq21}
DoF{{{ = }LM} \mathord{\left/
 {\vphantom { {(L + {1 \mathord{\left/
 {\vphantom {1 \varphi }} \right.
 \kern-\nulldelimiterspace} \varphi })}}} \right.
 \kern-\nulldelimiterspace} {(L + {1 \mathord{\left/
 {\vphantom {1 \varphi }} \right.
 \kern-\nulldelimiterspace} \varphi })}}.
\end{equation}
In practice, the number of cells $L$ is fixed and $\varphi $ is a function of $M$ and $N$, thus the value of DoF mainly depends on the relationship between $M$ and $N$. Define $\rho  = {M \mathord{\left/{\vphantom {M N}} \right.\kern-\nulldelimiterspace} N}$ as the transceiver antennas ratio  \cite{Maddah2012Completely} and then, the DoF can be simplified as
\begin{equation}
\label{eq22}
DoF{{{ = }LM} \mathord{\left/
 {\vphantom {{{ = }LM} {(L + {1 \mathord{\left/
 {\vphantom {1 {\left\lfloor {{1 \mathord{\left/
 {\vphantom {1 {(\rho  - 1)}}} \right.
 \kern-\nulldelimiterspace} {(\rho  - 1)}}} \right\rfloor }}} \right.
 \kern-\nulldelimiterspace} {\left\lfloor {{1 \mathord{\left/
 {\vphantom {1 {(\rho  - 1)}}} \right.
 \kern-\nulldelimiterspace} {(\rho  - 1)}}} \right\rfloor }})}}} \right.
 \kern-\nulldelimiterspace} {(L + {1 \mathord{\left/
 {\vphantom {1 {\left\lfloor {{1 \mathord{\left/
 {\vphantom {1 {(\rho  - 1)}}} \right.
 \kern-\nulldelimiterspace} {(\rho  - 1)}}} \right\rfloor }}} \right.
 \kern-\nulldelimiterspace} {\left\lfloor {{1 \mathord{\left/
 {\vphantom {1 {(\rho  - 1)}}} \right.
 \kern-\nulldelimiterspace} {(\rho  - 1)}}} \right\rfloor }})}}.
\end{equation}
To analyze the formula (22), we can get the conclusion that, with the increase of the ratio $\rho$, the DoF should be distinct. That is, when $M$ is not divisible by $ M -N $, the value of $\varphi = \left\lfloor {{N \mathord{\left/{\vphantom {N {\left( {M - N} \right)}}} \right.\kern-\nulldelimiterspace} {\left( {M - N} \right)}}} \right\rfloor $ will not change owing to the round down operation $\left\lfloor {*} \right\rfloor$ and the DoF keeps constant. While if $M$ is divisible by $ M -N $, the value of the DoF will grow sharply. Therefore, the value of $\varphi$ is divided into three cases, i.e., $\varphi > 2$, $\varphi=2$, $\varphi=1$, and the corresponding DoF of each situation is further discussed in Section V.

\subsection{Typical Application}
In this subsection, we take a typical application as example to demonstrate the whole process of the RIR scheme. In particular, the configuration of the network is set as $(L,M,K,N) = (3,4,2,3)$, i.e., the network is composed of three cells and in each cell, the base station with four antennas provides services to two users, where each user has three antennas. For this scenario, the details about the application of the RIR scheme are as follows.

The ﬁrst phase spans ${\cal L} = 3$ groups of $\varphi  = 3$ slots. From slot 1 to slot 3 of the first group, the symbols sent by the base station 1 are  as follows
\begin{equation}
\label{eq23}
\begin{array}{l}
{{\boldsymbol{S}}_1}[1] = \left( {\begin{array}{*{20}{c}}
{{a_1}},
{{a_2}},
{{b_1}},
{{b_2}}
\end{array}} \right)^T,\\
{{\boldsymbol{S}}_1}[2] = \left( {\begin{array}{*{20}{c}}
{{a_3}},
{{a_4}},
{{b_3}},
{{b_4}}
\end{array}} \right)^T, \\
{{\boldsymbol{S}}_1}[3] = \left( {\begin{array}{*{20}{c}}
{{a_5}},
{{a_6}},
{{b_5}},
{{b_6}}
\end{array}} \right)^T,
\end{array}
\end{equation}
where the sets ${\bf{A}} = \left\{ {{a_1}, \cdots ,{a_6}} \right\}$ and ${\bf{B}} = \left\{ {{b_1}, \cdots ,{b_6}} \right\}$ are the desired symbols for the users 1 and 2 in the target cell 1, respectively. Then, the received signals at user $k \in \left\{ {1,2} \right\}$ can be characterized as
\begin{equation}
\label{eq24}
\setlength{\arraycolsep}{2pt}
\begin{array}{l}
{{\bf{y}}_{1,k}}[t] = {\bf{H}}_1^{1,k}[t]{{\boldsymbol{S}}_1}[t]\\
\;\;\;\;\;\;\;\quad{   = }\left[ {\begin{array}{*{20}{c}}
{h_{1,1}^{k,1}[t]}&{h_{1,2}^{k,1}[t]}&{h_{1,3}^{k,1}[t]}&{h_{1,4}^{k,1}[t]}\\
{h_{1,1}^{k,2}[t]}&{h_{1,2}^{k,2}[t]}&{h_{1,3}^{k,2}[t]}&{h_{1,4}^{k,2}[t]}\\
{h_{1,1}^{k,3}[t]}&{h_{1,2}^{k,3}[t]}&{h_{1,3}^{k,3}[t]}&{h_{1,4}^{k,3}[t]}
\end{array}} \right]\left[ {\begin{array}{*{20}{c}}
{{a_{2t - 1}}}\\
{{a_{2t}}}\\
{{b_{2t - 1}}}\\
{{b_{2t}}}
\end{array}} \right]\\
\;\;\;\;\;\;\;\quad{   = }{{\boldsymbol{L}}_1}[t]\left( {{a_{2t - 1}},{a_{2t}},{b_{2t - 1}},{b_{2t}}} \right),t \in \{ 1,2,3\}.
\end{array}
\end{equation}
Since $M > N$, no users can decode the messages within one slot. From (24), we know that, in each slot, there are 4 independent values but we only have 3 equations. Thus, for this group, additional $\varphi  \times (M - N) = 3$ equations are needed. The users feed the channel estimation ${\bf{H}}_1^{1,k}[t]$ back to the base station 1. Similarly, in the second and third group, the transmitted symbols are defined as (25) and (26) respectively,
\begin{equation}
\label{eq25}
\begin{array}{l}
{{\boldsymbol{S}}_2}[4]\! =\! \left( {\begin{array}{*{20}{c}}
{{c_1}},
{{c_2}},
{{d_1}},
{{d_2}}
\end{array}} \right)^T,\\
{{\boldsymbol{S}}_2}[5]\! =\! \left( {\begin{array}{*{20}{c}}
{{c_3}},
{{c_4}},
{{d_3}},
{{d_4}}
\end{array}} \right)^T,\!\\
{{\boldsymbol{S}}_2}[6]\! = \!\left( {\begin{array}{*{20}{c}}
{{c_5}},
{{c_6}},
{{d_5}},
{{d_6}}
\end{array}} \right)^T,
\end{array}
\end{equation}
\begin{equation}
\label{eq26}
\begin{array}{l}
{{\boldsymbol{S}}_3}[7]\! =\! \left( {\begin{array}{*{20}{c}}
{{e_1}},
{{e_2}},
{{f_1}},
{{f_2}}
\end{array}} \right)^T,\\
{{\boldsymbol{S}}_3}[8]\! = \!\left( {\begin{array}{*{20}{c}}
{{e_3}},
{{e_4}},
{{f_3}},
{{f_4}}
\end{array}} \right)^T,\\
{{\boldsymbol{S}}_3}[9]\! =\! \left( {\begin{array}{*{20}{c}}
{{e_5}},
{{e_6}},
{{f_5}},
{{f_6}}
\end{array}} \right)^T.
\end{array}
\end{equation}
During the second group and the third group of the first phase, the base stations of the interference cells transmit signals to their served users in turn, but at the same time, the users in cell 1 will receive ICI inevitably, that is
\begin{equation}
\label{eq27}
\setlength{\arraycolsep}{2pt}
\begin{array}{l}
{{\bf{y}}_{1,k}}[t] = {\bf{H}}_2^{1,k}[t]{{\boldsymbol{S}}_2}[t]\\
\;\;\;\;\;\;\;\;\;\;\,{   = }\left[ {\begin{array}{*{20}{c}}
{h_{2,1}^{k,1}[t]}&{h_{2,2}^{k,1}[t]}&{h_{2,3}^{k,1}[t]}&{h_{2,4}^{k,1}[t]}\\
{h_{2,1}^{k,2}[t]}&{h_{2,2}^{k,2}[t]}&{h_{2,3}^{k,2}[t]}&{h_{2,4}^{k,2}[t]}\\
{h_{2,1}^{k,3}[t]}&{h_{2,2}^{k,3}[t]}&{h_{2,3}^{k,3}[t]}&{h_{2,4}^{k,3}[t]}
\end{array}} \right]\left[ {\begin{array}{*{20}{c}}
{{{c}_{2t - 7}}}\\
{{{c}_{2t - 6}}}\\
{{d_{2t - 7}}}\\
{{d_{2t - 6}}}
\end{array}} \right]\\
\;\;\;\;\;\;\;\;\;\;{   = }{{\textbf{L}}_2}\left( {{{c}_{2t - 7}},{{c}_{2t - 6}},{d_{2t - 7}},{d_{2t - 6}}} \right),t \in \{ 4,5,6\},
\end{array}
\end{equation}
\begin{equation}
\label{eq28}
\setlength{\arraycolsep}{2pt}
\begin{array}{l}
{{\bf{y}}_{1,k}}[t] = {\bf{H}}_3^{1,k}[t]{{\boldsymbol{S}}_3}[t]\\
\;\;\;\;\;\;\;\;\;\,{   = }\left[ {\begin{array}{*{20}{c}}
{h_{3,1}^{k,1}[t]}&{h_{3,2}^{k,1}[t]}&{h_{3,3}^{k,1}[t]}&{h_{3,4}^{k,1}[t]}\\
{h_{3,1}^{k,2}[t]}&{h_{3,2}^{k,2}[t]}&{h_{3,3}^{k,2}[t]}&{h_{3,4}^{k,2}[t]}\\
{h_{3,1}^{k,3}[t]}&{h_{3,2}^{k,3}[t]}&{h_{3,3}^{k,3}[t]}&{h_{3,4}^{k,3}[t]}
\end{array}} \right]\left[ {\begin{array}{*{20}{c}}
{{e_{2t - 13}}}\\
{{e_{2t{ - 12}}}}\\
{{f_{2t - 13}}}\\
{{f_{2t{ - }12}}}
\end{array}} \right]\\
\;\;\;\;\;\;\;\;\;{=}{{\textbf{L}}_3}\left( \!{{e_{2t - 13}},{e_{2t{ - 12}}},{f_{2t - 13}},{f_{2t{ - }12}}} \right)\!,\!t\! \in\! \{ 7,8,9\}
\end{array}
\end{equation}
From (27), we know that ${{\textbf{L}}_2}\left( {{{c}_{2t - 7}},{{c}_{2t - 6}}} \right),t \!\in\! \{4,5,6\} $ are the partical desired signals for the user 1 in cell 2 but the partical interference signals for the user 2 in cell 2, and ${{\textbf{L}}_2}\left( {{d_{2t - 7}},{d_{2t - 6}}} \right),t \!\in\! \{4,5,6\}$ is opposite to the former. In the meantime, the similar conclusion can be gotten from (28). Therefore, to utilize partial desired signals, partial interference signals should be eliminated with precoding matrix  which is desined at base station side. Therefore, the users feed interference signals ${{\textbf{L}}_2}\left( {{{c}_{2t - 7}},{{c}_{2t - 6}},{d_{2t - 7}},{d_{2t - 6}}} \right),t \!\in\! \{ 4,5,6\}$ and ${{\textbf{L}}_3}\left( {{e_{2t - 13}},{e_{2t{ - 12}}},{f_{2t - 13}},{f_{2t{ - }12}}} \right),t \!\in\! \{ 7,8,9\}$ back to the base station 1 to achieve interference elimination by ${{\bf{U}}_1}[t]$.

In the second phase, as the analysis before, \begin{small}$\left({{c}_{2t - 7}},{{c}_{2t - 6}}\right),t \!\in\! \{4,5,6\} $\end{small} and \begin{small}$\left( {{e_{2t - 13}},{e_{2t{ - }12}}} \right),t \!\in\! \{7,8,9\}$\end{small} are the desidered symbols for the user 1 of cell 2 and the user 1 of cell 3, respectively. After receiving feedback information, i.e., \begin{small}${\boldsymbol{y}}_{1,k}^{}[t],t \!\in\! \{4,5,6,7,8,9\} $\end{small}, the precoding matrix \begin{small}${{\bf{U}}_1}[t],t \!\in\! \{ 4,5,6\} $\end{small} and \begin{small}${{\bf{U}}_1}[t],t \!\in\! \{ 7,8,9\} $\end{small} are designed as (29) and (30), respectively.
\newcounter{mytempeqncnt2}
\begin{figure*}[!t]
\normalsize
\setcounter{mytempeqncnt2}{\value{equation}}
\setcounter{equation}{28}
\begin{equation}
\label{eq29}
\begin{array}{l}
{{\bf{U}}_1}[t]{{\textbf{L}}_2}[t] = {{\bf{U}}_1}[t]\left[ {\begin{array}{*{20}{c}}
{h_{2,1}^{k,1}[t]}&{h_{2,2}^{k,1}[t]}&{h_{2,3}^{k,1}[t]}&{h_{2,4}^{k,1}[t]}\\
{h_{2,1}^{k,2}[t]}&{h_{2,2}^{k,2}[t]}&{h_{2,3}^{k,2}[t]}&{h_{2,4}^{k,2}[t]}\\
{h_{2,1}^{k,3}[t]}&{h_{2,2}^{k,3}[t]}&{h_{2,3}^{k,3}[t]}&{h_{2,4}^{k,3}[t]}
\end{array}} \right]\left[ {\begin{array}{*{20}{c}}
{{{c}_{2t - 7}}}\\
{{{c}_{2t - 6}}}\\
{{d_{2t - 7}}}\\
{{d_{2t - 6}}}
\end{array}} \right]\\
\qquad\qquad\quad\!\! = {{\bf{U}}_1}[t]{\left[ {\begin{array}{*{20}{c}}
{h_{2,1}^{k,1}[t]}&{h_{2,1}^{k,2}[t]}&{h_{2,1}^{k,3}[t]}\\
{h_{2,2}^{k,1}[t]}&{h_{2,2}^{k,2}[t]}&{h_{2,2}^{k,3}[t]}
\end{array}} \right]^T}\left[ {\begin{array}{*{20}{c}}
{{{c}_{2t - 7}}}\\
{{{c}_{2t - 6}}}
\end{array}} \right]{ + }{{\bf{U}}_1}[t]{\left[ {\begin{array}{*{20}{c}}
{h_{2,3}^{k,1}[t]}&{h_{2,3}^{k,2}[t]}&{h_{2,3}^{k,3}[t]}\\
{h_{2,4}^{k,1}[t]}&{h_{2,4}^{k,2}[t]}&{h_{2,4}^{k,3}[t]}
\end{array}} \right]^T}\left[ {\begin{array}{*{20}{c}}
{{d_{2t - 7}}}\\
{{d_{2t - 6}}}
\end{array}} \right]\\
\qquad\qquad\quad\!\!= {{\bf{U}}_1}[t]{\left[ {\begin{array}{*{20}{c}}
{h_{2,1}^{k,1}[t]}&{h_{2,1}^{k,2}[t]}&{h_{2,1}^{k,3}[t]}\\
{h_{2,2}^{k,1}[t]}&{h_{2,2}^{k,2}[t]}&{h_{2,2}^{k,3}[t]}
\end{array}} \right]^T}\left[ {\begin{array}{*{20}{c}}
{{{c}_{2t - 7}}}\\
{{{c}_{2t - 6}}}
\end{array}} \right] + {\bf{0}}\\
\qquad\qquad\quad\!\!= {\bf{L}}_1^{2,1}({{c}_{2t - 7}},{{c}_{2t - 6}})[t],t \in \{ 4,5,6\} 
\end{array}
\end{equation}
\begin{equation}
\label{eq30}
{{\bf{U}}_1}[t]{_3}[t] = {\bf{L}}_1^{3,1}\left( {{e_{2t - 13}},{e_{2t{ - }12}}} \right)[t],t \in \{ 7,8,9\}.
\end{equation}
\hrulefill
\vspace*{4pt}
\end{figure*}
Note that for the user 2 in the interference cells, the partial interference signals ${\boldsymbol{L}}_1^{2,1}({{c}_{2t - 7}},{{c}_{2t - 6}})[t]$ and ${\boldsymbol{L}}_1^{3,1}\left( {{e_{2t - 13}},{e_{2t{ - }12}}} \right)[t]$ can also be used to eliminate IUI so that they can achieve equivalent functions for the user 2. Then the interference regeneration matrix is used at the base station 1 to avoid the additional interference in the third phases, i.e.
\begin{align}
\label{eq31-32}
&{{\bf{V}}_1}[t]{\boldsymbol{L}}_1^{2,1}({{c}_{2t - 7}},{{c}_{2t - 6}})[t]\nonumber\\
&={{\bf{V}}_3}[t]{\boldsymbol{L}}_3^{2,1}({{c}_{2t - 7}},{{c}_{2t - 6}})[t],\\
&{{\bf{V}}_1}[t]{\boldsymbol{L}}_1^{3,1}\left( {{e_{2t - 13}},{e_{2t{ - }12}}} \right)[t]\nonumber\\
&={{\bf{V}}_2}[t]{\boldsymbol{L}}_2^{3,1}\left( {{e_{2t - 13}},{e_{2t{ - }12}}} \right)[t].
\end{align} 
Herein, for the users in the cell 3, the additional partial interference signals ${{\bf{V}}_1}[t]\boldsymbol{L}_1^{2,1}({{c}_{2t - 7}},{{c}_{2t - 6}})[t]$ is convert to the partial interference signals, i.e., ${{\bf{V}}_3}[t]{\boldsymbol{L}}_3^{2,1}({{c}_{2t - 7}},{{c}_{2t - 6}})[t]$ that the later one is known by the users in the cell 3. Meanwhile, for the users in the cell 2, the signals are processed in the same way. Henceforth,  the additional interference caused by interference retransmission process can be avoid.

In the third phase, all the base stations simutanously retransmit partial desired signals as follows
\begin{align}
\label{eq33-35}
{{\bf{X}}_1}[10] &= \sum\limits_{t = 4}^6 {{{\bf{V}}_1}[t]} {\boldsymbol{L}}_1^{2,1}[t]{ + }\sum\limits_{t = 7}^9 {{{\bf{V}}_1}[t]} {\boldsymbol{L}}_1^{3,1}[t],\\
{{\bf{X}}_2}[10] &= \sum\limits_{t = 1}^3 {{{\bf{V}}_2}[t]} {\boldsymbol{L}}_2^{1,1}[t]{ + }\sum\limits_{t = 7}^9 {{{\bf{V}}_2}[t]} {\boldsymbol{L}}_2^{3,1}[t],\\
{{\bf{X}}_3}[10] &= \sum\limits_{t = 1}^3 {{{\bf{V}}_3}[t]} {\boldsymbol{L}}_3^{1,1}[t]{ + }\sum\limits_{t = 4}^6 {{{\bf{V}}_3}[t]} {\boldsymbol{L}}_3^{2,1}[t].
\end{align}
For the user $k$ in the cell 1, received signals can be written as
\begin{equation}
\label{eq36}
\begin{array}{l}
{{\bf{y}}_{1,k}}[10] = {\bf{H}}_1^{[1,k]}[10]\sum\limits_{t = 4}^6 {{{\bf{V}}_1}[t]\boldsymbol{L}_1^{2,1}[t]\left( {{c_{2t - 7}},{c_{2t - 6}}} \right)} \\
 + {\bf{H}}_1^{[1,k]}[10]\sum\limits_{t = 7}^9 {{{\bf{V}}_1}[t]{\boldsymbol{L}}_1^{3,1}[t]\left( {{e_{2t - 13}},{e_{2t - 12}}} \right)} \\
 + {\bf{H}}_2^{[1,k]}[10]\sum\limits_{t = 1}^3 {{{\bf{V}}_2}[t]{\boldsymbol{L}}_2^{1,1}[t]\left( {{a_{2t - 1}},{a_{2t}}} \right)} \\
 + {\bf{H}}_2^{[1,k]}[10]\sum\limits_{t = 7}^9 {{{\bf{V}}_2}[t]{\boldsymbol{L}}_2^{3,1}[t]\left( {{e_{2t - 13}},{e_{2t - 12}}} \right)} \\
 + {\bf{H}}_3^{[1,k]}[10]\sum\limits_{t = 1}^3 {{{\bf{V}}_3}[t]{\boldsymbol{L}}_3^{1,1}[t]\left( {{a_{2t - 1}},{a_{2t}}} \right)} \\
 + {\bf{H}}_3^{[1,k]}[10]\sum\limits_{t = 4}^6 {{{\bf{V}}_3}[t]{\boldsymbol{L}}_3^{2,1}[t]\left( {{c_{2t - 7}},{c_{2t - 6}}} \right)}.
\end{array}
\end{equation}

Owing to the existence of interference regeneration matrix, the partial interference signals can be converted to the partial interference signals known by the user $k$ in the cell 1. Thus, (36) can be simplified as
\begin{equation}
\label{eq37}
\begin{small}
\begin{array}{l}
{{\bf{y}}^{i,k}}[10]\;{ = }
\left( {{\bf{H}}_2^{[1,k]}[10]{ + }{\bf{H}}_3^{[1,k]}[10]} \right) \sum\limits_{t = 1}^3 {{{\bf{V}}_2}[t]{\boldsymbol{L}}_2^{1,1}[t]\left( {{a_{2t - 1}},{a_{2t}}} \right)} {}\\
{ + }\left( {{\bf{H}}_1^{[1,k]}[10]{ + }{\bf{H}}_3^{[1,k]}[10]} \right) \sum\limits_{t = 4}^6 {{{\bf{V}}_1}[t]{\boldsymbol{L}}_1^{2,1}[t]\left( {{{c}_{2t - 7}},{c_{2t{ - 6}}}} \right)} \\
{ + }\left( {{\bf{H}}_1^{[1,k]}[10]{ + }{\bf{H}}_2^{[1,k]}[10]} \right) \sum\limits_{t = 7}^9 {{{\bf{V}}_1}[t]{\boldsymbol{L}}_1^{3,1}[t]\left( {{e_{2t - 13}},{e_{2t{ - 12}}}} \right)}.
\end{array}
\end{small}
\end{equation}
For (37), we know that 
\begin{small}$({\bf{H}}_2^{[1,k]}[10]{ + }{\bf{H}}_3^{[1,k]}[10])\sum\limits_{t = 1}^3 {{{\bf{V}}_2}[t]{\boldsymbol{L}}_2^{1,1}[t]} \\({a_{2t - 1}},{a_{2t}})$\end{small} is the partial desired signals and the rest is the known partial interference signals. Therefore, for the user 1 in the target cell 1, the whole process can be summarized as the following equation (38) 
\newcounter{mytempeqncnt3}
\begin{figure*}[!t]
\normalsize
\setcounter{mytempeqncnt3}{\value{equation}}
\setcounter{equation}{37}
\begin{equation}
\label{eq38_1}
\hspace{-36.2em}
\left[ {\begin{array}{*{20}{c}}
{{{\boldsymbol{y}}_{1,1}}[1]},
{{{\boldsymbol{y}}_{1,1}}[2]},
{{{\boldsymbol{y}}_{1,1}}[3]},
{{{\boldsymbol{y}}_{1,1}}[10]}
\end{array}} \right]^T
{\rm{ = }}\nonumber
\end{equation}
\begin{small}
\begin{equation}
\label{eq38_2}
\hspace{-2mm}
\setlength{\arraycolsep}{0.5pt}
\left[ {\begin{array}{*{20}{c}}
{h_{1,1}^{1,1}[1]}&{h_{1,2}^{1,1}[1]}&{h_{1,3}^{1,1}[1]}&{h_{1,4}^{1,1}[1]}& \cdots &{}&{}&{}&{}\\
{h_{1,1}^{1,2}[1]}&{h_{1,2}^{1,2}[1]}&{h_{1,3}^{1,2}[1]}&{h_{1,4}^{1,2}[1]}& \cdots &{}&{}&{}&{}\\
{h_{1,1}^{1,3}[1]}&{h_{1,2}^{1,3}[1]}&{h_{1,3}^{1,3}[1]}&{h_{1,4}^{1,3}[1]}& \cdots &{}&{}&{}&{}\\
 \vdots & \vdots & \vdots & \vdots & \ddots & \vdots & \vdots & \vdots & \vdots \\
{}&{}&{}&{}& \cdots &{h_{1,1}^{1,1}[3]}&{h_{1,2}^{1,1}[3]}&{h_{1,3}^{1,1}[3]}&{h_{1,4}^{1,1}[3]}\\
{}&{}&{}&{}& \cdots &{h_{1,1}^{1,2}[3]}&{h_{1,2}^{1,2}[3]}&{h_{1,3}^{1,2}[3]}&{h_{1,4}^{1,2}[3]}\\
{}&{}&{}&{}& \cdots &{h_{1,1}^{1,3}[3]}&{h_{1,2}^{1,3}[3]}&{h_{1,3}^{1,3}[3]}&{h_{1,4}^{1,3}[3]}\\
{\sum\limits_{j = 2,3} {h_{j,1}^{1,1}[10]} {\boldsymbol\beta _1}}&{\sum\limits_{j = 2,3} {h_{j,2}^{1,1}[10]} {\boldsymbol\beta _1}}&{\sum\limits_{j = 2,3} {h_{j,3}^{1,1}[10]} {\boldsymbol\beta _1}}&{\sum\limits_{j = 2,3} {h_{j,4}^{1,1}[10]} {\boldsymbol\beta _1}}& \cdots &{\sum\limits_{j = 2,3} {h_{j,1}^{1,1}[10]} {\boldsymbol\beta _3}}&{\sum\limits_{j = 2,3} {h_{j,2}^{1,1}[10]} {\boldsymbol\beta _3}}&{\sum\limits_{j = 2,3} {h_{j,3}^{1,1}[10]} {\boldsymbol\beta _3}}&{\sum\limits_{j = 2,3} {h_{j,4}^{1,1}[10]} {\boldsymbol\beta _3}}\\
{\sum\limits_{j = 2,3} {h_{j,1}^{1,2}[10]} {\boldsymbol\beta _1}}&{\sum\limits_{j = 2,3} {h_{j,2}^{1,2}[10]} {\boldsymbol\beta _1}}&{\sum\limits_{j = 2,3} {h_{j,3}^{1,2}[10]} {\boldsymbol\beta _1}}&{\sum\limits_{j = 2,3} {h_{j,4}^{1,2}[10]} {\boldsymbol\beta _1}}& \cdots &{\sum\limits_{j = 2,3} {h_{j,1}^{1,2}[10]} {\boldsymbol\beta _3}}&{\sum\limits_{j = 2,3} {h_{j,2}^{1,2}[10]} {\boldsymbol\beta _3}}&{\sum\limits_{j = 2,3} {h_{j,3}^{1,2}[10]} {\boldsymbol\beta _3}}&{\sum\limits_{j = 2,3} {h_{j,4}^{1,2}[10]} {\boldsymbol\beta _3}}\\
{\sum\limits_{j = 2,3} {h_{j,1}^{1,3}[10]} {\boldsymbol\beta _1}}&{\sum\limits_{j = 2,3} {h_{j,2}^{1,3}[10]} {\boldsymbol\beta _1}}&{\sum\limits_{j = 2,3} {h_{j,3}^{1,3}[10]} {\boldsymbol\beta _1}}&{\sum\limits_{j = 2,3} {h_{j,4}^{1,3}[10]} {\boldsymbol\beta _1}}& \cdots &{\sum\limits_{j = 2,3} {h_{j,1}^{1,3}[10]} {\boldsymbol\beta _3}}&{\sum\limits_{j = 2,3} {h_{j,2}^{1,3}[10]} {\boldsymbol\beta _3}}&{\sum\limits_{j = 2,3} {h_{j,3}^{1,3}[10]} {\boldsymbol\beta _3}}&{\sum\limits_{j = 2,3} {h_{j,4}^{1,3}[10]} {\boldsymbol\beta _3}}
\end{array}} \right]
\!\!\!\!  \left[ {\begin{array}{*{20}{c}}
{{a_1}}\\
{{a_2}}\\
{{b_1}}\\
{{b_2}}\\
 \vdots \\
{{a_5}}\\
{{a_6}}\\
{{b_5}}\\
{{b_6}}
\end{array}} \right]
\end{equation}
\end{small}
\hrulefill
\vspace*{4pt}
\end{figure*}
where ${{\boldsymbol{\beta}} _t}{ = }{v_2}[t]{u_2}[t]h_1^{2,1}[t] + {v_3}[t]{u_3}[t]h_1^{3,1}[t]$. It is obvious that the channel matrix is a $12 \times 12$ non-singular matrix. Herein, user 1 in the target cell 1 can decode its messages with delayed CSI.

From the above illustration, 36 independent symbols are transmitted by three base stations over 10 coherent slots. Therefore, the DoF of the proposed RIR scheme is 3.6. While under the same conditions, i.e., $(L,M,K,N) = (3,4,2,3)$, the DoF for the TDMA scheme and RIA scheme are only 3 and 3.28, respectively.

\section{Beamforming Based Distributed Retrospective Interference Alignment Scheme}
For the proposed RIR scheme, though it can achieve greater DoF than the RIA scheme, the scheme has the problem of performance degradation as the transceiver antennas ratio $\rho$ approaches 1. To handle this issue, a new beamforming based distributed retrospective interference alignment (B-DRIA) scheme is proposed. The core idea of this scheme is that the beamforming technique is adopted to eliminate ICI and then, several additional slots are taken for the base stations to simultaneously transmit desired signals. Finally, we design distributed retrospective interference matrices to align the IUI at the user side and decode the messages separately. In the following, we firstly illustrate the details of the proposed scheme which includes three phases, i.e., CSIT acquisition process, cellular beamforming and signal transmission, and distributed retrospective interference alignment. Then, the DoF of the B-DRIA scheme is analyzed. To make it understandable, we also present an example of the scheme at last.

\subsection{CSIT Acquisition Process}
The first phase takes $L$ slots and in which the base stations take turns to send messages to its served users. Since $M > N$, these desired messages cannot be decoded within only one slot and additional transmissions are required. Therefore, the users do not complete decoding in the current slot but feedback the estimated CSI to the base stations. The CSI  contains both the channel gain estimation and the DoA estimation where the former part is used for retrospective interference alignment and the latter part is taken in cellular  beamforming. Taking the cell 1 for example, the process of the first phase is illustrated as Fig. 7.
\begin{figure}[!t]
\centering
\includegraphics[width=2.5in]{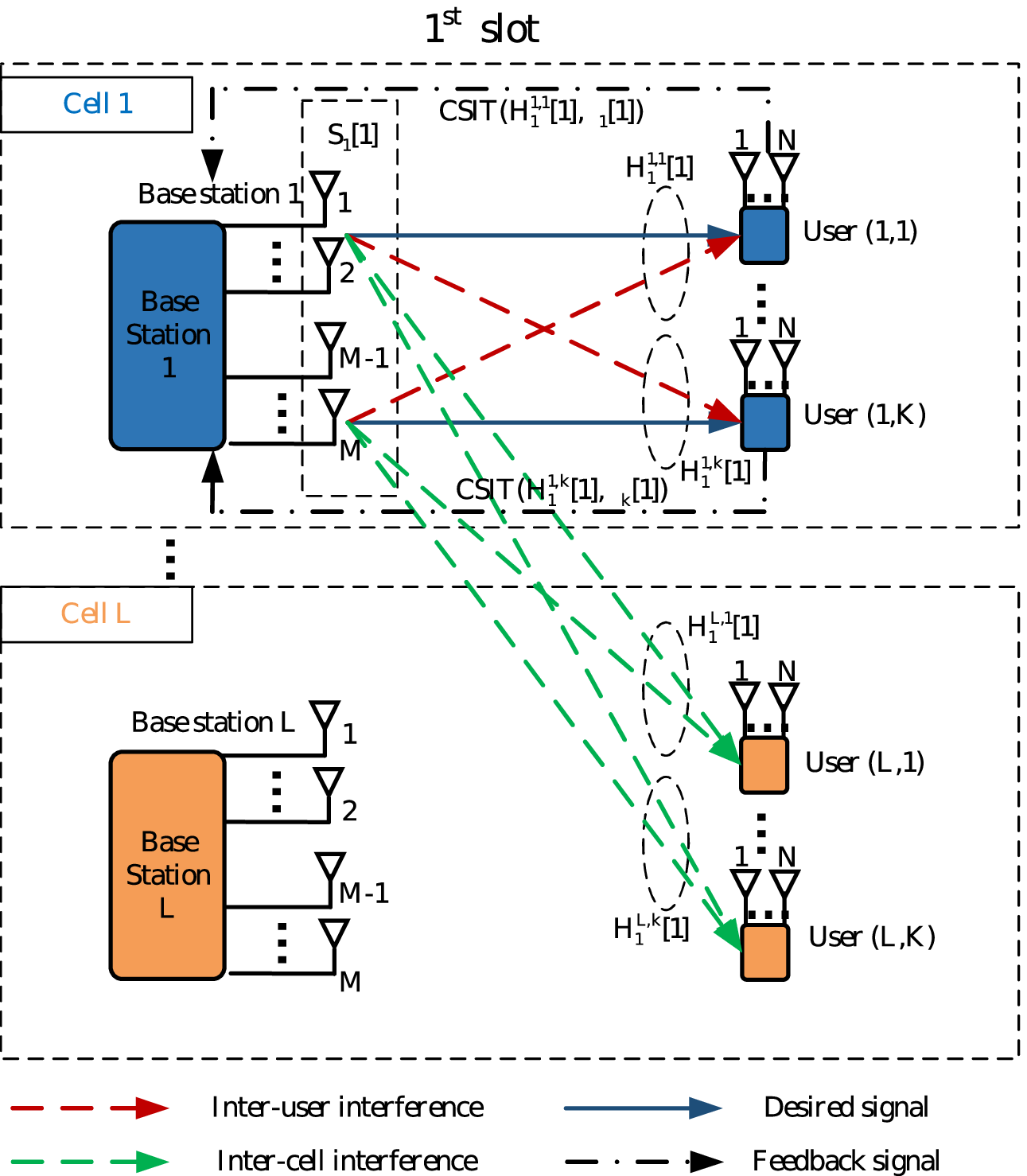}
\caption{CSIT Acquisition process in 1st phase.}
\label{Fig7}
\end{figure}
To make it universal applicable, assume that in the slot $i \in {\cal L}$, the base station $i \in {\cal L}$ transmits messages as
\begin{equation}
\label{eq39}
{\boldsymbol{S}}_{i}^{}[i] = \left[ {{s_{i,1}}[i], \cdots {s_{i,\left\lceil {{M \mathord{\left/
 {\vphantom {M K}} \right.
 \kern-\nulldelimiterspace} K}} \right\rceil }}[i], \cdots ,{s_{i,K\left\lceil {{M \mathord{\left/
 {\vphantom {M K}} \right.
 \kern-\nulldelimiterspace} K}} \right\rceil }}[i]} \right].
\end{equation}
Then for user $k$ in the target cell $j$, its received signals can be expressed in time domain and angular domain as follows
\begin{align}
\label{eq40-41}
&{\boldsymbol{y}}_{j,k}^{}[i] = {\bf{H}}_i^{j,k}[t]{{\boldsymbol{S}}_i}[t],i \in {\cal L},j = i,\\
&{\boldsymbol{y}}_{j,k}^{}[t] = {\bf{A}}_i^{j,k}[t]{{\boldsymbol{S}}_i}[t],i \in {\cal L},j = i.
\end{align} 
Based on the received signals, the LS algorithm \cite{2004Discussion} and the multiple signal classification algorithm algorithm (MUSIC) \cite{6422415} can be used at the user side to estimate the channel gain and DoA, respectively. With channel estimation, the feedback information from user $k$ is extended into two parts, i.e., the channel gain information and the angel information, and its form can be characterized by
\begin{equation}
\label{eq42}
CSIT{ \!=\! }\left\{ {\left( {{\bf{H}}_i^{i,1}[i]\!,\!{\theta _1}[i]} \right)\!,\! \!\cdots\! \!,\!\left( {{\bf{H}}_i^{i,k}[i]\!,\!{\theta _k}[i]} \right)} \right\}\!,\!i \in {\cal L},
\end{equation}
where $\left( {{\bf{H}}_i^{i,k}[i],{\theta _k}[i]} \right)$ denotes the channel matrix from base station $i$ to user $k$ in the target cell $i$ and the corresponding DoA of that user.

\subsection{Cellular Beamforming and Signal Transmission}
Following the CSIT acquisition process, we can further perform the process of cellular beamforming and signal transmission. While this phase spans $\bar \varphi  = \left\lfloor {{{2N - M} \mathord{\left/{\vphantom {{2N - M} {M - N}}} \right.\kern-\nulldelimiterspace} {M - N}}} \right\rfloor $ slots, the base stations simultaneously transmit independent messages to their served users. However, the desired signals received at the user sides are contaminated by both ICI and IUI. Therefore, additional operations are needed to eliminate  interference signals.

In order to cancel the ICI, beamforming is adopted at the base stations. Define the slots set of this phase as $\Psi { = }\left\{ {L + 1,\cdots ,L{ + }\bar \varphi } \right\}$ and the beamforming matrices are
\begin{align}
\label{eq43-44}
&{\bf{W}}_i^{j,k}[t]{ = }{{\bf{0}}_{M \times M}},j \in {\cal L}\backslash i,k \in {\textbf{K}},t \in \Psi,\\
&{\bf{W}}_i^{j,k}[t]{ = }{{\bf{I}}_{M \times M}},i = j,k \in {\textbf{K}},t \in \Psi.
\end{align} 
where ${{\bf{0}}_{M \times M}}$ is a zero matrix and ${{\bf{I}}_{M \times M}}$ is an $M$-dimensional identity matrix. Then, for one slot $t,\forall t \in \Psi $, the signals sent by base station $i$ to user $k$ in the target cell $j$ is 
\begin{equation}
\begin{array}{l}
\label{eq45}
{\boldsymbol{X}}_i^{j,k}[t] = {\bf{W}}_i^{j,k}[t]{\boldsymbol{S}}_{i}^{}[t]\\
\qquad\quad\, = {{\bf{I}}_{M \times M}}{\left[ {{s_{i,1}}[t], \cdots ,{s_{i,M}}[t]} \right]^T}\\
\qquad\quad\, = {\left[ {{s_{i,1}}[t], \cdots ,{s_{i,M}}[t]} \right]^T}.
\end{array}
\end{equation}
After performing the beamforming with ${\bf{W}}_i^{j,k}$, the ICI is eliminated which means that all the users will not receive signals from the base stations of the interference cells. This process is shown in Fig. 8.
\begin{figure}[!t]
\centering
\includegraphics[width=2.5in]{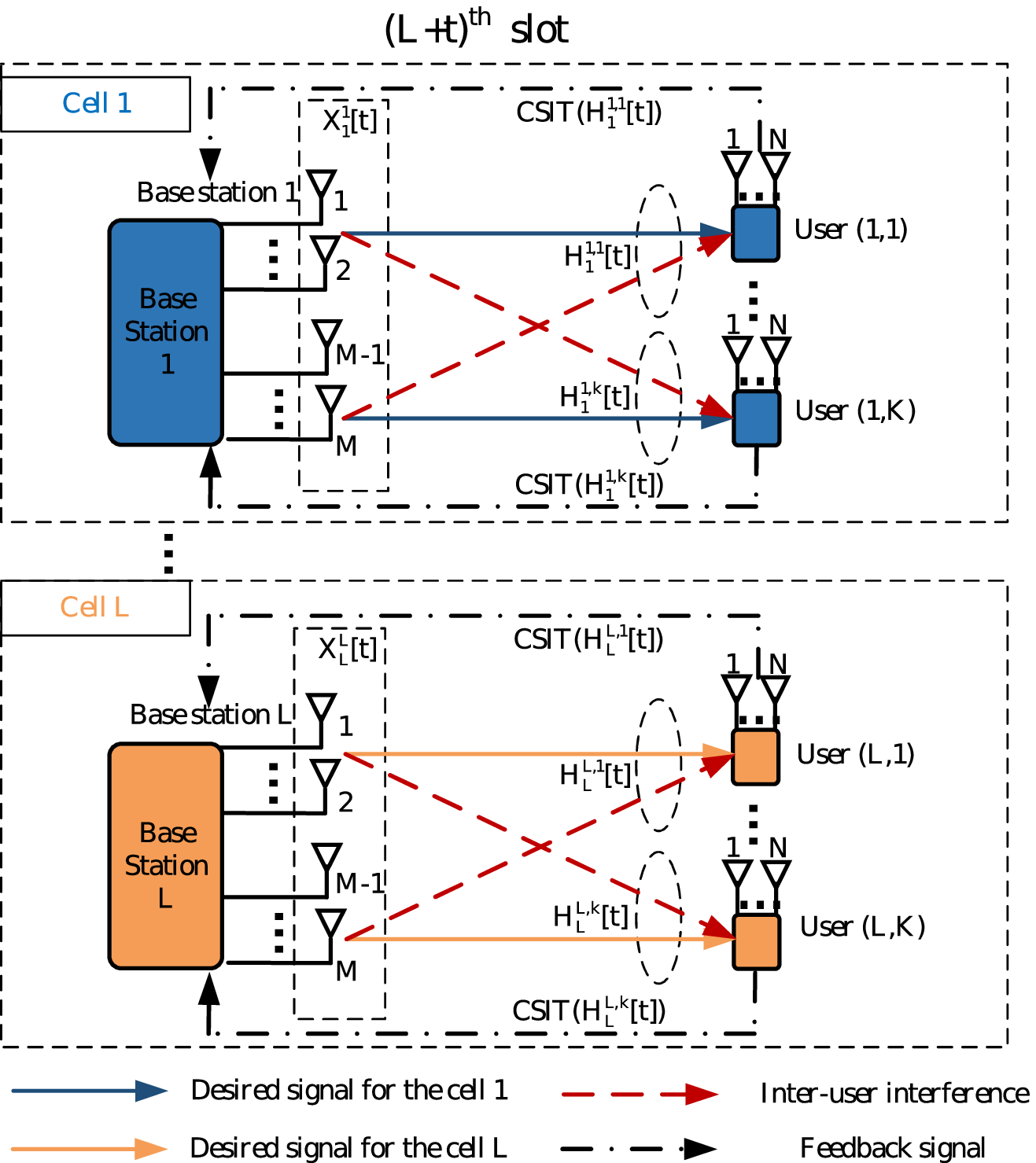}
\caption{Signals transmission with beamforming in 2nd phase.}
\label{Fig8}
\end{figure}
Thus, the received signals of the user $k$ in the cell $j$ can be expressed as 
\begin{equation}
\label{eq46}
{\boldsymbol{y}}_{j,k}^{}[t] = {\bf{H}}_i^{i,k}[t]{\boldsymbol{X}}_i^{j,k}[t].
\end{equation}
Due to the presence of the IUI, no users can decode the received message in one slot. By analyzing the components of the received signals, we find that, for user $k$, the IUI contained desired information for the other users, i.e., the received signals of the user $k$ can be written as
\begin{equation}
\begin{array}{l} 
\label{eq47}
{\boldsymbol{y}}_{j,k}^{}[t] = {\bf{H}}_i^{i,k}[t]{{\boldsymbol{S}}_i}[t] + {\bf{H}}_i^{i, - k}[t]{{\boldsymbol{S}}_i}[t]\\
\qquad\quad\!= \boldsymbol{L}\left( {{s_{i,(k - 1)\left\lceil {{M \mathord{\left/
 {\vphantom {M K}} \right.
 \kern-\nulldelimiterspace} K}} \right\rceil  + 1}}[L + t], \cdots ,{s_{i,k\left\lceil {{M \mathord{\left/
 {\vphantom {M K}} \right.
 \kern-\nulldelimiterspace} K}} \right\rceil }}[L + t]} \right)\\
\qquad\quad\! + \boldsymbol{I} \left( {{s_{i,(k - 1)\left\lceil {{M \mathord{\left/
 {\vphantom {M K}} \right.
 \kern-\nulldelimiterspace} K}} \right\rceil  + 1}}[L + t], \cdots ,{s_{i,k\left\lceil {{M \mathord{\left/
 {\vphantom {M K}} \right.
 \kern-\nulldelimiterspace} K}} \right\rceil }}[L + t]} \right)\\
\qquad\quad\!= \boldsymbol{L}_i^{i,k}[t]{ + }\boldsymbol{I}_i^{i,k}[t], - k = {\textbf{K}}\backslash k,
\end{array}
\end{equation}
where $\boldsymbol{L}_i^{i,k}[t]$ represents the desired signals and $ \boldsymbol{I}_i^{i,k}[t]$ represents IUI. Therefore, through $L{ + }\bar \varphi$ slots, there are $\bar \varphi M + M$ messages needed to be decoded for each user and we only have $\bar \varphi N + N$ equations. In order to decode these messages, some extra information is desired, i.e., to formulate $N$ additional equations of each user.

\subsection{Distributed Retrospective Interference Alignment}
Based on the first phase and the second phase, especially the CSI estimation and feedback, each base station can obtain the knowledge of the CSI and the IUI feedback, i.e., $ \boldsymbol{I}_i^{i,k}[t]$. Therefore, the distributed retrospective interference alignment is developed to make use of the combination of the IUI from the previous slots. Part of which can be extracted by the users to supplement $N$ additional equations. In specific, the designed  vector from the base station in the cell $i$ is shown as follows
\begin{equation}
\setlength{\arraycolsep}{2pt}
\label{eq48}
{{\boldsymbol{S}}_i}[L\! + \!\bar \varphi {\! +\! }1]
{ = }\left[ {\begin{array}{*{20}{c}}
{\sum\limits_{k = 1}^K {\left( { I_i^{i,k}[i]\! + \! I_i^{i,k}[L\! + \!1] \!+ \! \cdots \! + \! I_i^{i,k}[L\! + \!\bar \varphi ]} \right)} }\\
0\\
 \vdots \\
0
\end{array}} \right],
\end{equation}
where ${\sum\limits_{k = 1}^K {\left( { \boldsymbol{I}_i^{i,k}[i]\! + \! \boldsymbol{I}_i^{i,k}[L\! + \!1] \!+ \! \cdots \! + \! \boldsymbol{I}_i^{i,k}[L\! + \!\bar \varphi ]} \right)} } $ is a $N$-dimensional vector composed of the whole past slots’ IUI. By the pre-mentioned beamforming process and the distributed interference alignment scheme, the base stations are capable of transmitting the vector ${{\boldsymbol{S}}_i}[L\! + \!\bar \varphi {\! +\! }1]$ without introducing new ICI or IUI, as shown in Fig. 9.
\begin{figure}[!t]
\centering
\includegraphics[width=2.5in]{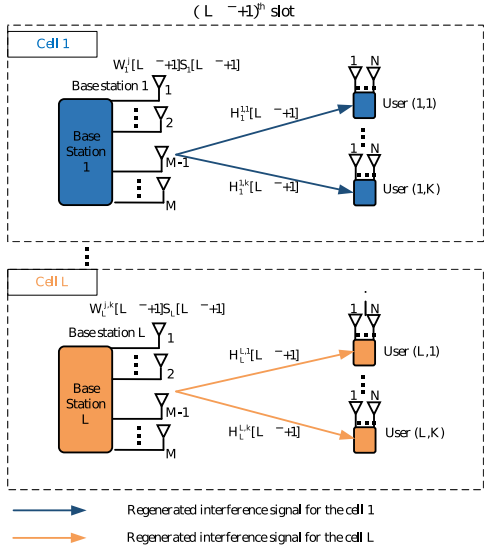}
\caption{Regenerated interference retransmission in 3rd phase.}
\label{Fig9}
\end{figure}
Then, the received signals of the user $k$ in the cell $j$ can be expressed as
\begin{equation}
\label{eq49}
\begin{array}{l}
{\boldsymbol{y}}_{j,k}^{}[L + \bar \varphi { + }1]\\
= {\bf{H}}_i^{j,k}[L + \bar \varphi { + }1]{\bf{W}}_i^{j,k}[L + \bar \varphi { + }1]{{\boldsymbol{S}}_i}[L + \bar \varphi { + }1].
\end{array}
\end{equation}
From (49), the desired signals can be extracted from the combination of the IUI by the interference elimination process
\begin{equation}
\begin{array}{l}
\label{eq50}
{\boldsymbol{y}}_{j,k}^{}[L + \bar \varphi { + }1] - {\bf{H}}_i^{j,k}[L + \bar \varphi { + }1]\sum\limits_{t = 1}^{L + \bar \varphi } { \boldsymbol{I}_i^{i, - k}[t]}\\
= {\bf{H}}_i^{j,k}[L + \bar \varphi { + }1]( \boldsymbol{I}_i^{i,k}[i] + \sum\limits_{t = L + 1}^{L + \bar \varphi } { \boldsymbol{I}_i^{i,k}[t]} ),
\end{array}
\end{equation}
where $\boldsymbol{I}_i^{i,k}[i] + \sum\limits_{t = L + 1}^{L + \bar \varphi } { \boldsymbol{I}_i^{i,k}[t]}$ is the supplementary of the $N$ additional equations. Herein, the whole process of the proposed B-DRIA scheme can be summarized in (51), 
\newcounter{mytempeqncnt4}
\begin{figure*}[!t]
\normalsize
\setcounter{mytempeqncnt4}{\value{equation}}
\setcounter{equation}{50} 
\begin{equation}
\hspace{-2mm}
\label{eq50}
\underbrace {\left[ {\begin{array}{*{20}{c}}
{{\bf{y}}_{j,k}^{}[i]}\\
{{\bf{y}}_{j,k}^{}[L + 1]}\\
 \vdots \\
{{\bf{y}}_{j,k}^{}[L + \bar \varphi ]}\\
{{\bf{y}}_{j,k}^{}[L + \bar \varphi  + 1]}
\end{array}} \right]}_{\bf{Y}}{ = }\underbrace {\left[ {\begin{array}{*{20}{c}}
{{\bf{H}}_i^{i,k}[i]}&0& \cdots &0\\
0&{{\bf{H}}_i^{i,k}[L + 1]}& \cdots & \vdots \\
 \vdots & \ddots & \ddots &0\\
0& \cdots &0&{{\bf{H}}_i^{i,k}[L + \bar \varphi ]}\\
{{\bf{H}}_i^{i,k}[L + \bar \varphi { + }1]{{\bf{\beta }}_i}}&{{\bf{H}}_i^{i,k}[L + \bar \varphi { + }1]{{\bf{\beta }}_{L + 1}}}& \cdots &{{\bf{H}}_i^{i,k}[L + \bar \varphi { + }1]{{\bf{\beta }}_{L + \bar \varphi }}}
\end{array}} \right]}_{\bf{H}}\underbrace {\left[ {\begin{array}{*{20}{c}}
{{{\bf{S}}_i}[i]}\\
{{{\bf{S}}_i}[L + 1]}\\
 \vdots \\
{{{\bf{S}}_i}[L + \bar \varphi ]}
\end{array}} \right]}_{\bf{S}}
\end{equation}
\hrulefill
\vspace*{4pt}
\end{figure*}
where ${{\boldsymbol{\beta}} _t}$ is the precoding matrix defined as follows
\begin{equation}
\label{eq52}
{{\boldsymbol{\beta}} _t} = {\bf{W}}_i^{j,k}[t]\sum\nolimits_{k ={\textbf{K}} \backslash k} {{\bf{H}}_i^{i,k}[t]}.
\end{equation}
Since the delayed CSIT is adopted in the transmission process, which means that, for any two different slots, the time span exceeds the coherence time, the channels are independent with each other over different slots, e.g., ${\bf{H}}_i^{i,k}[L + 1]$ and ${\bf{H}}_i^{i,k}[L + \bar \varphi ]$. In addition, in each slot, the channel matrix is non-singular, which means that each submatrix has a full rank. Herein, the matrix $\bf{H}$ is a $\left\lfloor {{N \mathord{\left/
 {\vphantom {N {M - N}}} \right.\kern-\nulldelimiterspace} {M - N}}} \right\rfloor M$ dimensional non-singular matrix, and the LS method \cite{2004Discussion} can be used to decode the desired messages.

\subsection{Degrees of Freedom}
Following the above illustration, we analyze the DoF of the proposed B-DRIA scheme. Since the base stations spend $L{ + }\bar \varphi { + }1$  incoherent slots and transmit $L\left\lceil {{M \mathord{\left/ {\vphantom {M K}} \right.\kern-\nulldelimiterspace} K}} \right\rceil K(\bar \varphi { + }1)$  independent symbols, we assume that the transmission is reliable, i.e., the value of signal-to-noise ratio (SNR) is approximated to infinity. Hence the DoF of the B-DRIA scheme can be calculated as
\begin{equation}
\label{eq53}
DoF{ = }{{L\left\lceil {{M \mathord{\left/
 {\vphantom {M K}} \right.
 \kern-\nulldelimiterspace} K}} \right\rceil K(\bar \varphi { + }1)} \mathord{\left/
 {\vphantom {{L\left\lceil {{M \mathord{\left/
 {\vphantom {M K}} \right.
 \kern-\nulldelimiterspace} K}} \right\rceil K(\bar \varphi { + }1)} {(L{ + }\bar \varphi { + }1)}}} \right.
 \kern-\nulldelimiterspace} {(L{ + }\bar \varphi { + }1)}}.
\end{equation}
To facilitate the analysis, we assume that $M$ is divisible by $K$. Then, the above equation can be simplified as
\begin{equation}
\label{eq54}
\begin{array}{l}
DoF{{{ = }LM} \mathord{\left/
 {\vphantom {{{ = }LM} {(1 + {L \mathord{\left/
 {\vphantom {L {(\bar \varphi { + }1)}}} \right.
 \kern-\nulldelimiterspace} {(\bar \varphi { + }1)}})}}} \right.
 \kern-\nulldelimiterspace} {(1 + {L \mathord{\left/
 {\vphantom {L {(\bar \varphi { + }1)}}} \right.
 \kern-\nulldelimiterspace} {(\bar \varphi { + }1)}})}}.
\end{array}
\end{equation}
As mentioned earlier, the number of cells $L$ is constant and the signal transmission slots in the second phase $\bar \varphi$ is a function of $M$ and $N$. Hence, the value of DoF mainly depends on transceiver $\rho$ and $M$, and the DoF is rewritten as
\begin{equation}
\label{eq55}
DoF{{{ = }LM} \mathord{\left/
 {\vphantom {{{ = }LM} {(1 + {L \mathord{\left/
 {\vphantom {L {\left\lfloor {{3 \mathord{\left/
 {\vphantom {3 {(\rho  - 1)}}} \right.
 \kern-\nulldelimiterspace} {(\rho  - 1)}}} \right\rfloor }}} \right.
 \kern-\nulldelimiterspace} {\left\lfloor {{3 \mathord{\left/
 {\vphantom {3 {(\rho  - 1)}}} \right.
 \kern-\nulldelimiterspace} {(\rho  - 1)}}} \right\rfloor }})}}} \right.
 \kern-\nulldelimiterspace} {(1 + {L \mathord{\left/
 {\vphantom {L {\left\lfloor {{3 \mathord{\left/
 {\vphantom {3 {(\rho  - 1)}}} \right.
 \kern-\nulldelimiterspace} {(\rho  - 1)}}} \right\rfloor }}} \right.
 \kern-\nulldelimiterspace} {\left\lfloor {{3 \mathord{\left/
 {\vphantom {3 {(\rho  - 1)}}} \right.
 \kern-\nulldelimiterspace} {(\rho  - 1)}}} \right\rfloor }})}}.
\end{equation}
With the increase of $\rho$, the valued of the DoF can be separately discussed under two cases/conditions. That is, when $M$ is not divisible by $ M -N $, the value of $\bar \varphi = \left\lfloor {{{2N - M} \mathord{\left/{\vphantom {{2N - M} {M - N}}} \right.\kern-\nulldelimiterspace} {M - N}}} \right\rfloor$ will not change owing to the round down operation $\left\lfloor {*} \right\rfloor$ and the DoF keeps constant. While if $M$ is divisible by $ M -N $, the value of the DoF will grow sharply. Therefore, the value of $\bar \varphi $ is divided into three cases, i.e., $\bar \varphi > 1$, $\bar \varphi =1$, $\bar \varphi =0$, and the corresponding DoF gains of each situation are further discussed in Section V.

\subsection{Typical Application}
For the comparison purpose, the condition used herein is the same as that adopted for the RIR scheme, that is, $(L,M,K,N) = (3,4,2,3)$, which means that the network is composed of 3 cells and in each cell, the base station with 4 antennas provides services to 2 users, where each user has 3 antennas. For this scenario, the B-DRIA scheme is adopted and the whole process is illustrated as follows.

The ﬁrst phase takes 3 slots, in which the base stations take turns to send messages. In each slot, the symbols sent by the base stations are shown as follows
\begin{equation}
\label{eq56}
\begin{array}{l}
{{\boldsymbol{S}}_1}[1] = \left( {\begin{array}{*{20}{c}}
{{a_1}},
{{a_2}},
{{b_1}},
{{b_2}}
\end{array}} \right)^T,\\
{{\boldsymbol{S}}_2}[2] = \left( {\begin{array}{*{20}{c}}
{{c_1}},
{{c_2}},
{{d_1}},
{{d_2}}
\end{array}} \right)^T,\\
{{\boldsymbol{S}}_3}[3] = \left( {\begin{array}{*{20}{c}}
{{e_1}},
{{e_2}},
{{f_1}},
{{f_2}}
\end{array}} \right)^T.
\end{array}
\end{equation}
Take the user $k$ in the target cell 1 for example, in the slot 1, the received signals can be expressed as
\begin{equation}
\label{eq57}
\setlength{\arraycolsep}{2pt}
\begin{array}{l}
{\boldsymbol{y}}_{1,k}^{}[1]\! =\!  {\bf{H}}_1^{1,k}[1]{{\boldsymbol{S}}_1}[1]\\  
\qquad\quad\!\! =\!  \left[ {\begin{array}{*{20}{c}}
{h_{1,1}^{k,1}[1]}&{h_{1,2}^{k,1}[1]}&{h_{1,3}^{k,1}[1]}&{h_{1,4}^{k,1}[1]}\\
{h_{1,1}^{k,2}[1]}&{h_{1,2}^{k,2}[1]}&{h_{1,3}^{k,2}[1]}&{h_{1,4}^{k,2}[1]}\\
{h_{1,1}^{k,3}[1]}&{h_{1,2}^{k,3}[1]}&{h_{1,3}^{k,3}[1]}&{h_{1,4}^{k,3}[1]}
\end{array}} \right]\left[ {\begin{array}{*{20}{c}}
{{a_1}}\\
{{a_2}}\\
{{b_1}}\\
{{b_2}}
\end{array}} \right]
\end{array},
\end{equation}
\begin{equation}
\label{eq58}
\setlength{\arraycolsep}{2pt}
\begin{array}{l}
{\boldsymbol{y}}_{1,k}^{}[1]\!  = \! {\bf{A}}_1^{1,k}[1]{{\boldsymbol{S}}_1}[1]\\
\qquad\quad\!\! = \! \left[ {\begin{array}{*{20}{c}}
1&1&1&1\\
{{e^{ - j{\mu _1}}}}&{{e^{ - j{\mu _2}}}}&{{e^{ - j{\mu _3}}}}&{{e^{ - j{\mu _4}}}}\\
{{e^{ - j2{\mu _1}}}}&{{e^{ - j2{\mu _2}}}}&{{e^{ - j2{\mu _3}}}}&{{e^{ - j2{\mu _4}}}}
\end{array}} \right]\left[ {\begin{array}{*{20}{c}}
{{a_1}}\\
{{a_2}}\\
{{b_1}}\\
{{b_2}}
\end{array}} \right]
\end{array}.
\end{equation}
The former is the time domain expression where the channel gain matrix is estimated by the LS algorithm\cite{2004Discussion}, and the latter is the angular domain expression where the DoA matrix is obtained by the MUSIC algorithm \cite{6422415}. Therefore, the feedback estimation contains two parts, i.e., the channel gain information ${\bf{H}}_1^{1,k}[1]$ and the angel information ${\theta _1}[1]$. Since $M > N$, user $k$ cannot decode the messages ${{\bf{S}}_1}[1]$ within one slot and meanwhile, the estimated CSI is fed back to the base station 1. Similarly, the user $k$ of cell 2 receive the signals in slot 2 and the user $k$ of cell 3 receive signals in slot 3 as
\begin{equation}
\label{eq59}
\setlength{\arraycolsep}{2pt}
\begin{array}{l}
{\bf{y}}_{2,k}^{}[2] = {\bf{H}}_2^{2,k}[2]{{\boldsymbol{S}}_2}[2]\\
 \;\;\;\;\;\;\;\;= \left[ {\begin{array}{*{20}{c}}
{h_{2,1}^{k,1}[2]}&{h_{2,2}^{k,1}[2]}&{h_{2,3}^{k,1}[2]}&{h_{2,4}^{k,1}[2]}\\
{h_{2,1}^{k,2}[2]}&{h_{2,2}^{k,2}[2]}&{h_{2,3}^{k,2}[2]}&{h_{2,4}^{k,2}[2]}\\
{h_{2,1}^{k,3}[2]}&{h_{2,2}^{k,3}[2]}&{h_{2,3}^{k,3}[2]}&{h_{2,4}^{k,3}[2]}
\end{array}} \right]\left[ {\begin{array}{*{20}{c}}
{{c_1}}\\
{{c_2}}\\
{{d_1}}\\
{{d_2}}
\end{array}} \right]
\end{array},
\end{equation}
\vspace{-0.5em}
\begin{equation}
\setlength{\arraycolsep}{2pt}
\label{eq60}
\begin{array}{l}
{\bf{y}}_{3,k}^{}[t] = {\bf{H}}_3^{3,k}[t]{{\boldsymbol{S}}_3}[t]\\
\;\;\;\;\;\;\;\;{   = }\left[ {\begin{array}{*{20}{c}}
{h_{3,1}^{k,1}[3]}&{h_{3,2}^{k,1}[3]}&{h_{3,3}^{k,1}[3]}&{h_{3,4}^{k,1}[3]}\\
{h_{3,1}^{k,2}[3]}&{h_{3,2}^{k,2}[3]}&{h_{3,3}^{k,2}[3]}&{h_{3,4}^{k,2}[3]}\\
{h_{3,1}^{k,3}[3]}&{h_{3,2}^{k,3}[3]}&{h_{3,3}^{k,3}[3]}&{h_{3,4}^{k,3}[3]}
\end{array}} \right]\left[ {\begin{array}{*{20}{c}}
{{e_1}}\\
{{e_2}}\\
{{f_1}}\\
{{f_2}}
\end{array}} \right],
\end{array}
\end{equation}
By channel estimation, both channel gain information ${\bf{H}}_2^{2,k}[2]$ and angel information ${\theta _2}[2]$ are obtained and fed back to the base station 2, and it is the same for ${\bf{H}}_3^{3,k}[3]$ and ${\theta _3}[3]$.

In the second phase, after obtaining CSI, the beamforming is used by the base station to eliminate the ICI. Taking base station 1  for example,  beamforming matrices are written as
\begin{equation}
\label{eq61}
\begin{array}{l}
{\bf{W}}_1^{j,k}[t]{ = }\left[ {\begin{array}{*{20}{c}}
0&0&0&0\\
0&0&0&0\\
0&0&0&0\\
0&0&0&0
\end{array}} \right]\\
{ = }{{\sl\bf{0}}_{4 \times 4}},j \in {\cal L}\backslash 1,t \in \Psi ,
\end{array}
\end{equation}
\begin{equation}
\label{eq62}
\begin{array}{l}
{\bf{W}}_1^{1,k}[t]{ = }\left[ {\begin{array}{*{20}{c}}
1&0&0&0\\
0&1&0&0\\
0&0&1&0\\
0&0&0&1
\end{array}} \right]\\
{ = }{{\bf{I}}_{4 \times 4}},t \in \Psi ,
\end{array}
\end{equation}
where $\Psi$ denotes the slots set of the second phase. For base station 1, the transmitted signals are changed to
\begin{equation}
\label{eq63}
\begin{array}{l}
{{\bf{X}}_1}[t] = {{\bf{W}}_1}[t]{{\boldsymbol{S}}_1}[t]\\
 = {\left[ {\begin{array}{*{20}{c}}
{{{\bf{I}}_{4 \times 4}}}&{{{\bf{0}}_{4 \times 4}}}&{{{\bf{0}}_{4 \times 4}}}
\end{array}} \right]^T}{\left[ {\begin{array}{*{20}{c}}
{{a_1}}&{{a_2}}&{{b_1}}&{{b_2}}
\end{array}} \right]^T}.
\end{array}
\end{equation}
After the process of the beamforming, the base stations are able to simultaneously transmit signals without introducing additional ICI. However, the unsolvable issue which caused by the condition $M > N$ is still exist. While from (59) and (60), it is not difficult to infer that, for the case  $M=4$ and $N =3$, only 1 ($=M - N$) additional equation is desired in each slot. In order to make full use of the time resources, 2 more slots are taken to transmit signals, i.e., $\Psi { = }\left\{ {4,5} \right\}$. Herein, for user $k$ in cell 1, in slot 4 and slot 5, the received signals can be written as follows, respectively,
\begin{equation}
\label{eq64}
\begin{small}
\begin{array}{l}
{\bf{y}}_{1,k}^{}[4] \\
={\bf{H}}_1^{1,k}[4]{\bf{W}}_1^{1,k}[4]{{\boldsymbol{S}}_1}[4]\\
= \left[ {\begin{array}{*{20}{c}}
{h_{1,1}^{k,1}[4]}&{h_{1,2}^{k,1}[4]}&{h_{1,3}^{k,1}[4]}&{h_{1,4}^{k,1}[4]}\\
{h_{1,1}^{k,2}[4]}&{h_{1,2}^{k,2}[4]}&{h_{1,3}^{k,2}[4]}&{h_{1,4}^{k,2}[4]}\\
{h_{1,1}^{k,3}[4]}&{h_{1,2}^{k,3}[4]}&{h_{1,3}^{k,3}[4]}&{h_{1,4}^{k,3}[4]}
\end{array}} \right]\left[ {\begin{array}{*{20}{c}}
{{a_3}}\\
{{a_4}}\\
{{b_3}}\\
{{b_4}}
\end{array}} \right]\\
=\left[ {\begin{array}{*{20}{c}}
{h_{1,1}^{k,1}[4]}&{h_{1,2}^{k,1}[4]}\\
{h_{1,1}^{k,2}[4]}&{h_{1,2}^{k,2}[4]}\\
{h_{1,1}^{k,3}[4]}&{h_{1,2}^{k,3}[4]}
\end{array}} \right]\left[ {\begin{array}{*{20}{c}}
{{a_3}}\\
{{a_4}}
\end{array}} \right]\vspace{0.5ex}\\
+\left[ {\begin{array}{*{20}{c}}
{h_{1,3}^{k,1}[4]}&{h_{1,4}^{k,1}[4]}\\
{h_{1,3}^{k,2}[4]}&{h_{1,4}^{k,2}[4]}\\
{h_{1,3}^{k,3}[4]}&{h_{1,4}^{k,3}[4]}
\end{array}} \right]\left[ {\begin{array}{*{20}{c}}
{{b_3}}\\
{{b_4}}
\end{array}} \right]\\
={\bf{L}}_1^{1,k}[4]{ + }{\bf{I}}_1^{1,k}[4],
\end{array}
\end{small}
\end{equation}
\begin{equation}
\label{eq65}
\begin{small}
\begin{array}{l}
{\bf{y}}_{1,k}^{}[5] \\
={\bf{H}}_1^{1,k}[5]{\bf{W}}_1^{1,k}[5]{{\boldsymbol{S}}_1}[5]\\
= \left[ {\begin{array}{*{20}{c}}
{h_{1,1}^{k,1}[5]}&{h_{1,2}^{k,1}[5]}&{h_{1,3}^{k,1}[5]}&{h_{1,4}^{k,1}[5]}\\
{h_{1,1}^{k,2}[5]}&{h_{1,2}^{k,2}[5]}&{h_{1,3}^{k,2}[5]}&{h_{1,4}^{k,2}[5]}\\
{h_{1,1}^{k,3}[5]}&{h_{1,2}^{k,3}[5]}&{h_{1,3}^{k,3}[5]}&{h_{1,4}^{k,3}[5]}
\end{array}} \right]\left[ {\begin{array}{*{20}{c}}
{{a_5}}\\
{{a_6}}\\
{{b_5}}\\
{{b_6}}\\
\end{array}} \right]\\
=\left[ {\begin{array}{*{20}{c}}
{h_{1,1}^{k,1}[5]}&{h_{1,2}^{k,1}[5]}\\
{h_{1,1}^{k,2}[5]}&{h_{1,2}^{k,2}[5]}\\
{h_{1,1}^{k,3}[5]}&{h_{1,2}^{k,3}[5]}
\end{array}} \right]\left[ {\begin{array}{*{20}{c}}
{{a_5}}\\
{{a_6}}\\
\end{array}} \right] \vspace{0.5ex} \\
+\left[ {\begin{array}{*{20}{c}}
{h_{1,3}^{k,1}[5]}&{h_{1,4}^{k,1}[5]}\\
{h_{1,3}^{k,2}[5]}&{h_{1,4}^{k,2}[5]}\\
{h_{1,3}^{k,3}[5]}&{h_{1,4}^{k,3}[5]}
\end{array}} \right]\left[ {\begin{array}{*{20}{c}}
{{b_5}}\\
{{b_6}}\\
\end{array}} \right]\\
= {\boldsymbol{L}}_1^{1,k}[5]{ + }{\boldsymbol{I}}_1^{1,k}[5].
\end{array}
\end{small}
\end{equation}
Note that $\boldsymbol{L}_i^{i,k}[t]$ represents the desired signals and $\boldsymbol{I}_i^{i,k}[t]$ represents the IUI and which means that, if $k=1$, the former is the desired signals while the later is the IUI, that is
\begin{equation}
\label{eq66}
{\boldsymbol{L}}_1^{1,1}[5]{ = }\left[ {\begin{array}{*{20}{c}}
{h_{1,1}^{1,1}[5]}&{h_{1,2}^{1,1}[5]}\\
{h_{1,1}^{1,2}[5]}&{h_{1,2}^{1,2}[5]}\\
{h_{1,1}^{1,3}[5]}&{h_{1,2}^{1,3}[5]}
\end{array}} \right]\left[ {\begin{array}{*{20}{c}}
{{a_5}}\\
{{a_6}}
\end{array}} \right],
\end{equation}
\begin{equation}
\vspace{-0.5em}
\label{eq67}
{\boldsymbol{I}}_1^{1,1}[5]{ = }\left[ {\begin{array}{*{20}{c}}
{h_{1,3}^{1,1}[5]}&{h_{1,4}^{1,1}[5]}\\
{h_{1,3}^{1,2}[5]}&{h_{1,4}^{1,2}[5]}\\
{h_{1,3}^{1,3}[5]}&{h_{1,4}^{1,3}[5]}
\end{array}} \right]\left[ {\begin{array}{*{20}{c}}
{{b_5}}\\
{{b_6}}
\end{array}} \right],
\end{equation}
and this is the same for the case $k=2$. Now, the signals from different cells are independent, thus they are able to align the interference separately and the DRIA algorithm is adopted.

In the third phase, with the knowledge of delayed CSIT, base station 1 regenerates interference signals of  previous slots as
\begin{small}
\begin{equation}
\label{eq68}
\setlength{\arraycolsep}{2pt}
\begin{array}{l}
{{\boldsymbol{ I}}_1^{1,1}}[1] + {{\boldsymbol{ I}}_1^{1,2}}[1]\\
\!=\! \left[ {\begin{array}{*{20}{c}}
{h_{1,3}^{1,1}[1]}&{h_{1,4}^{1,1}[1]}\\
{h_{1,3}^{1,2}[1]}&{h_{1,4}^{1,2}[1]}\\
{h_{1,3}^{1,3}[1]}&{h_{1,4}^{1,3}[1]}
\end{array}} \right]\left[ {\begin{array}{*{20}{c}}
{{b_1}}\\
{{b_2}}
\end{array}} \right]\!+\! \left[ {\begin{array}{*{20}{c}}
{h_{1,3}^{2,1}[1]}&{h_{1,4}^{2,1}[1]}\\
{h_{1,3}^{2,2}[1]}&{h_{1,4}^{2,2}[1]}\\
{h_{1,3}^{2,3}[1]}&{h_{1,4}^{2,3}[1]}
\end{array}} \right]\left[ {\begin{array}{*{20}{c}}
{{a_1}}\\
{{a_2}}
\end{array}} \right]\\
\!=\! \left[ {\begin{array}{*{20}{c}}
{{{\boldsymbol{h}}_{1,3}^2}[1]}&{{{\boldsymbol{h}}_{1,4}^2}[1]}&{{{\boldsymbol{h}}_{1,3}^1}[1]}&{{{\boldsymbol{h}}_{1,4}^1[1]}}
\end{array}} \right]\left[ {\begin{array}{*{20}{c}}
{{a_1}}\\
{{a_2}}\\
{{b_1}}\\
{{b_2}}
\end{array}} \right],
\end{array}
\end{equation}
\vspace{-0.5em}
\begin{equation}
\label{eq69}
\setlength{\arraycolsep}{2pt}
\begin{array}{l}
{{\boldsymbol{I}}_1^{1,1}}[4] + {{\boldsymbol{I}}_1^{1,2}}[4] \\
\!=\! \left[ {\begin{array}{*{20}{c}}
{h_{1,3}^{1,1}[4]}&{h_{1,4}^{1,1}[4]}\\
{h_{1,3}^{1,2}[4]}&{h_{1,4}^{1,2}[4]}\\
{h_{1,3}^{1,3}[4]}&{h_{1,4}^{1,3}[4]}
\end{array}} \right]\left[ {\begin{array}{*{20}{c}}
{{b_3}}\\
{{b_4}}
\end{array}} \right]\!+\!\left[ {\begin{array}{*{20}{c}}
{h_{1,3}^{2,1}[4]}&{h_{1,4}^{2,1}[4]}\\
{h_{1,3}^{2,2}[4]}&{h_{1,4}^{2,2}[4]}\\
{h_{1,3}^{2,3}[4]}&{h_{1,4}^{2,3}[4]}
\end{array}} \right]\left[ {\begin{array}{*{20}{c}}
{{a_3}}\\
{{a_4}}
\end{array}} \right]\\
\!= \!\left[ {\begin{array}{*{20}{c}}
{{{\boldsymbol{h}}_{1,3}^2}[4]}&{{{\boldsymbol{h}}_{1,4}^2}[4]}&{{{\boldsymbol{h}}_{1,3}^1}[4]}&{{{\boldsymbol{h}}_{1,4}^1}[4]}
\end{array}} \right]\left[ {\begin{array}{*{20}{c}}
{{a_3}}\\
{{a_4}}\\
{{b_3}}\\
{{b_4}}
\end{array}} \right],
\end{array}
\end{equation}
\begin{equation}
\vspace{-0.5em}
\label{eq70}
\setlength{\arraycolsep}{2pt}
\begin{array}{l}
\boldsymbol{I_1^{1,1}}[5] +\boldsymbol{ I_1^{1,2}}[5]\\ 
\!=\! \left[ {\begin{array}{*{20}{c}}
{h_{1,3}^{1,1}[5]}&{h_{1,4}^{1,1}[5]}\\
{h_{1,3}^{1,2}[5]}&{h_{1,4}^{1,2}[5]}\\
{h_{1,3}^{1,3}[5]}&{h_{1,4}^{1,3}[5]}
\end{array}} \right]\left[ {\begin{array}{*{20}{c}}
{{b_5}}\\
{{b_6}}
\end{array}} \right]\!+ \!\left[ {\begin{array}{*{20}{c}}
{h_{1,3}^{2,1}[5]}&{h_{1,4}^{2,1}[5]}\\
{h_{1,3}^{2,2}[5]}&{h_{1,4}^{2,2}[5]}\\
{h_{1,3}^{2,3}[5]}&{h_{1,4}^{2,3}[5]}
\end{array}} \right]\left[ {\begin{array}{*{20}{c}}
{{a_5}}\\
{{a_6}}
\end{array}} \right]\\
\!= \!\left[ {\begin{array}{*{20}{c}}
{{{\boldsymbol{h}}_{1,3}^2}[5]}&{{{\boldsymbol{h}}_{1,4}^2}[5]}&{{{\boldsymbol{h}}_{1,3}^1}[5]}&{{{\boldsymbol{h}}_{1,4}^1}[5]}
\end{array}} \right]\left[ {\begin{array}{*{20}{c}}
{{a_5}}\\
{{a_6}}\\
{{b_5}}\\
{{b_6}}
\end{array}} \right],
\end{array}
\end{equation}
\end{small}
where ${\boldsymbol{h}}_{i,M}^j[t]$ denotes the channel vector from antenna $M$ of the base station $i$ to user $k$ in the t-th slot and it has the form as below
\begin{equation}
\label{eq71}
{\boldsymbol{h}}_{i,M}^j[t]{ = }{\left[ {\begin{array}{*{20}{c}}
{h_{i,M}^{j,1}[t]}&{h_{i,M}^{j,2}[t]}&{h_{i,M}^{j,3}[t]}
\end{array}} \right]^T}.
\end{equation}
Then the transmitted symbols can be characterized as
\begin{small}
\begin{equation}\
\hspace{-1em}
\label{eq72}
\setlength{\arraycolsep}{2pt}
\begin{array}{l}
{{\boldsymbol{S}}_1}[6]{ \!=\! } \\
\left[ {\begin{array}{*{20}{c}}
{({\boldsymbol{I}}_1^{1,1}[1]\! + \! {\boldsymbol{I}}_1^{1,2}[1])\! +\! ( {\boldsymbol{I}}_1^{1,1}[4]\! + \! {\boldsymbol{I}}_1^{1,2}[4])\! + \!( {\boldsymbol{I}}_1^{1,1}[5] \!+\!  {\boldsymbol{I}}_1^{1,2}[5])}\\
{{0_{1 \times 1}}}
\end{array}} \right].
\end{array}
\end{equation}
\end{small}
\!\!For user 1 in the target cell 1, the received signals are
\begin{equation}
\label{eq73}
\begin{array}{l}
{{\boldsymbol{y}}_{1,1}}[6] = {\bf{H}}_1^{1,1}[6]{\bf{W}}_1^{1,1}[6]{{\boldsymbol{S}}_1}[6]\\
\quad\quad\quad = \left[ {\begin{array}{*{20}{c}}
{h_{1,1}^{1,1}[6]}&{h_{1,2}^{1,1}[6]}&{h_{1,3}^{1,1}[6]}\\
{h_{1,1}^{1,2}[6]}&{h_{1,2}^{1,2}[6]}&{h_{1,3}^{1,2}[6]}\\
{h_{1,1}^{1,3}[6]}&{h_{1,2}^{1,3}[6]}&{h_{1,3}^{1,3}[6]}
\end{array}} \right]{{\boldsymbol{S}}_1}[6]\\
\quad\quad\quad= \left[ {\begin{array}{*{20}{c}}
{\boldsymbol{h}_1^{1,1}[6]}\\
{\boldsymbol{h}_1^{1,2}[6]}\\
{\boldsymbol{h}_1^{1,3}[6]}
\end{array}} \right]{{\boldsymbol{S}}_1}[6],
\end{array}
\end{equation}
where ${\boldsymbol{h}}_i^{k,N}[t]$ denotes the channel vector which is consist of the channel gains from the whole antennas of the base station $i$ to the $N$th antenna of the user $k$, and the vector is given by
\begin{equation}
\label{eq74}
{\boldsymbol{h}}_i^{k,N}[t]{ = }\left[ {\begin{array}{*{20}{c}}
{h_{i,1}^{k,N}[t]}&{h_{i,2}^{k,N}[t]}&{h_{i,3}^{k,N}[t]}
\end{array}} \right].
\end{equation}
Now, we have $(\bar \varphi  + 1) \times (M - N) = 3$ additional equations and for user 1 in the target cell 1, the whole process can be summarized by the equation (75). 
\newcounter{mytempeqncnt5}
\begin{figure*}[!t]
\normalsize
\setcounter{mytempeqncnt5}{\value{equation}}
\setcounter{equation}{74}
\begin{small}
\begin{equation}
\label{75}
\begin{array}{l}
\left[ {\begin{array}{*{20}{c}}
{{{\bf{y}}_{1,1}}[1]}\\
{{{\bf{y}}_{1,1}}[4]}\\
{{{\bf{y}}_{1,1}}[5]}\\
{{{\bf{y}}_{1,1}}[6]}
\end{array}} \right]{\rm{ = }}\left[ {\begin{array}{*{20}{c}}
{{{\bf{A}}_{3 \times 4}}}&{{{\bf{0}}_{3 \times 4}}}&{{{\bf{0}}_{3 \times 4}}}\\
{{{\bf{0}}_{3 \times 4}}}&{{{\bf{B}}_{3 \times 4}}}&{{{\bf{0}}_{3 \times 4}}}\\
{{{\bf{0}}_{3 \times 4}}}&{{{\bf{0}}_{3 \times 4}}}&{{{\bf{C}}_{3 \times 4}}}\\
{{{\bf{D}}_{3 \times 4}}}&{{{\bf{E}}_{3 \times 4}}}&{{{\bf{F}}_{3 \times 4}}}
\end{array}} \right]\left[ {\begin{array}{*{20}{c}}
{{{\bf{X}}_{4 \times 1}}}\\
{{{\bf{Y}}_{4 \times 1}}}\\
{{{\bf{Z}}_{4 \times 1}}}
\end{array}} \right] \quad where\\
{\rm{  }}{{\bf{A}}_{3 \times 4}} = \left[ {\begin{array}{*{20}{c}}
{h_{1,1}^{1,1}[1]}&{h_{1,2}^{1,1}[1]}&{h_{1,3}^{1,1}[1]}&{h_{1,4}^{1,1}[1]}\\
{h_{1,1}^{1,2}[1]}&{h_{1,2}^{1,2}[1]}&{h_{1,3}^{1,2}[1]}&{h_{1,4}^{1,2}[1]}\\
{h_{1,1}^{1,3}[1]}&{h_{1,2}^{1,3}[1]}&{h_{1,3}^{1,3}[1]}&{h_{1,4}^{1,3}[1]}
\end{array}} \right],{{\bf{B}}_{3 \times 4}} = \left[ {\begin{array}{*{20}{c}}
{h_{1,1}^{1,1}[4]}&{h_{1,2}^{1,1}[4]}&{h_{1,3}^{1,1}[4]}&{h_{1,4}^{1,1}[4]}\\
{h_{1,1}^{1,2}[4]}&{h_{1,2}^{1,2}[4]}&{h_{1,3}^{1,2}[4]}&{h_{1,4}^{1,2}[4]}\\
{h_{1,1}^{1,3}[4]}&{h_{1,2}^{1,3}[4]}&{h_{1,3}^{1,3}[4]}&{h_{1,4}^{1,3}[4]}
\end{array}} \right],\\
{{\bf{C}}_{3 \times 4}} = \left[ {\begin{array}{*{20}{c}}
{h_{1,1}^{1,1}[5]}&{h_{1,2}^{1,1}[5]}&{h_{1,3}^{1,1}[5]}&{h_{1,4}^{1,1}[5]}\\
{h_{1,1}^{1,2}[5]}&{h_{1,2}^{1,2}[5]}&{h_{1,3}^{1,2}[5]}&{h_{1,4}^{1,2}[5]}\\
{h_{1,1}^{1,3}[5]}&{h_{1,2}^{1,3}[5]}&{h_{1,3}^{1,3}[5]}&{h_{1,4}^{1,3}[5]}
\end{array}} \right],
{\rm{           }}{{\bf{X}}_{4 \times 1}} = \left[ {\begin{array}{*{20}{c}}
{{a_1}}\\
{{a_2}}\\
{{b_1}}\\
{{b_2}}
\end{array}} \right],{{\bf{Y}}_{4 \times 1}} = \left[ {\begin{array}{*{20}{c}}
{{a_3}}\\
{{a_4}}\\
{{b_3}}\\
{{b_4}}
\end{array}} \right],{{\bf{Z}}_{4 \times 1}} = \left[ {\begin{array}{*{20}{c}}
{{a_5}}\\
{{a_6}}\\
{{b_5}}\\
{{b_6}}
\end{array}} \right],\\
{\rm{           }}{{\bf{D}}_{3 \times 4}} = \left[ {\begin{array}{*{20}{c}}
{h_{1,3}^2[1]h_1^{1,1}[6]}&{h_{1,4}^2[1]h_1^{1,1}[6]}&{h_{1,3}^1[1]h_1^{1,1}[6]}&{h_{1,4}^1[1]h_1^{1,1}[6]}\\
{h_{1,3}^2[1]h_1^{1,2}[6]}&{h_{1,4}^2[1]h_1^{1,2}[6]}&{h_{1,3}^1[1]h_1^{1,2}[6]}&{h_{1,4}^1[1]h_1^{1,2}[6]}\\
{h_{1,3}^2[1]h_1^{1,3}[6]}&{h_{1,4}^2[1]h_1^{1,3}[6]}&{h_{1,3}^1[1]h_1^{1,3}[6]}&{h_{1,4}^1[1]h_1^{1,3}[6]}
\end{array}} \right],\\
{\rm{          }}{{\bf{E}}_{3 \times 4}} = \left[ {\begin{array}{*{20}{c}}
{h_{1,3}^2[4]h_1^{1,1}[6]}&{h_{1,4}^2[4]h_1^{1,1}[6]}&{h_{1,3}^1[4]h_1^{1,1}[6]}&{h_{1,4}^1[4]h_1^{1,1}[6]}\\
{h_{1,3}^2[4]h_1^{1,2}[6]}&{h_{1,4}^2[4]h_1^{1,2}[6]}&{h_{1,3}^1[4]h_1^{1,2}[6]}&{h_{1,4}^1[4]h_1^{1,2}[6]}\\
{h_{1,3}^2[4]h_1^{1,3}[6]}&{h_{1,4}^2[4]h_1^{1,3}[6]}&{h_{1,3}^1[4]h_1^{1,3}[6]}&{h_{1,4}^1[4]h_1^{1,3}[6]}
\end{array}} \right],\\
{\rm{           }}{{\bf{F}}_{3 \times 4}} = \left[ {\begin{array}{*{20}{c}}
{h_{1,3}^2[5]h_1^{1,1}[6]}&{h_{1,4}^2[5]h_1^{1,1}[6]}&{h_{1,3}^1[5]h_1^{1,1}[6]}&{h_{1,4}^1[5]h_1^{1,1}[6]}\\
{h_{1,3}^2[5]h_1^{1,2}[6]}&{h_{1,4}^2[5]h_1^{1,2}[6]}&{h_{1,3}^1[5]h_1^{1,2}[6]}&{h_{1,4}^1[5]h_1^{1,2}[6]}\\
{h_{1,3}^2[5]h_1^{1,3}[6]}&{h_{1,4}^2[5]h_1^{1,3}[6]}&{h_{1,3}^1[5]h_1^{1,3}[6]}&{h_{1,4}^1[5]h_1^{1,3}[6]}
\end{array}} \right].
\end{array}
\end{equation}
\end{small}
\hrulefill
\vspace*{4pt}
\end{figure*}
It is obvious that the channel matrix $\bf{H}$ is a $12 \times 12$ non-singular matrix and 36 independent symbols can be decoded over 6 slots. Thus, the DoF of the B-DRIA is 6, which is better than both the RIA scheme and the RIR scheme.

\section{Numerical Results}
In this section, numerical results are presented to characterize the performance of the proposed two IA schemes. In specific, to analyze the performance of the proposed schemes, two benchmark schemes are introduced:
\begin{itemize}
\vspace{-0.3em}
\item The TDMA scheme \cite{2006Ameliorated}, that is at each slot, only one base station is selected to transmit signals while the other base stations keep silent so that the negative effect of the ICI is cancelled. However, owing to the pattern of single base station transmission, both space resources and time resources are wasted inevitably. Hence, the obtained DoF is relaxed to the outer bound \cite{5074376}.

\item The RIA scheme \cite{7065324}, that is in the previous $t$ slots, only one base station is selected to transmit signals while the other base stations keep silent, and in the last slot, all the base stations make use of the combination of IUI in the previous slots to achieve simultaneous transmission. Therefore, the RIA scheme gets DoF gain while the improvement is still limited. The main reason is that the scheme only improves  the space and time resources' utilization in the last slot.
\vspace{-0.3em}
\end{itemize}
Herein, we develop the RIR scheme and B-DRIA scheme to further improve the utilization of time and space resources. As mentioned in Section III and Section IV, in the specific scenario, both RIR scheme and B-DRIA scheme achieve DoF gain better than RIR scheme, but the degree of the improvement of the DoF mainly depends on the relationship between transceiver antennas ratio $\rho$ and transmitter antennas $M$, whereas the relationship is nonlinear correlated. Therefore, the relationship should be further discussed in two cases:
\begin{itemize}
\vspace{-0.3em}
\item Fix the value of $\rho$ and change the value of $M$, and for this case, the interval of $\rho$ is from 1 to 2. In specific, owing to the assumption that no users can decode the messages with only one-slot transmitted signals, the lower bound of $\rho$ should be greater than 1. Meanwhile, in order to make comparison with the RIA scheme, the upper bound of $\rho$ should be no more than 2. Therefore, we fix the value of $\rho$ at 1, 3/2 or 2 and change the value of $N$ from 5 to 25. The effects of $M$ on the DoF are shown in Fig. 10, Fig. 11 and Fig. 12.

\item Fix the value of $M$ and change the value of $N$, and for this case, when $N$ increases and $M$ can be divided by $M-N$, the DoF grows sharply, otherwise the value of DoF keeps constant. Therefore, we fix the value of $M$ at 72 to make it divisible by as many integers as possible, and then change the value of $N$ from 36 to 72. The effects of $\rho$ on the DoF are shown in Fig. 13 and Fig. 14.
\vspace{-0.5em}
\end{itemize}

\subsection{Performance Analysis with Fixed $\rho$}
At first, we analyze how does the DoF is affected by the parameter $M$ under the condition that the parameter $\rho$ is fixed, i.e., $\rho  = {3 \mathord{\left/
 {\vphantom {3 2}} \right. \kern-\nulldelimiterspace} 2}$. In specific, the configuration of the network is set to $[L,M,K,N = (2,\left\lfloor {{3 \mathord{\left/ {\vphantom {3 2}} \right.\kern-\nulldelimiterspace} 2}N} \right\rfloor ,3,N)$, and meanwhile the value of the $N$ is varying from 5 to 25. Under this configuration, for the RIR scheme, $\varphi  = 2$ slots per group are used for signal transmission and at the same time, for the B-DRIA scheme, $\bar \varphi  = 1$ slot is used for signal transmission, respectively. From the result shown in Fig. 10, we observe that, for all the value of the $M$, B-DRIA scheme always has the best DoF performance, then it is the RIR scheme, and the RIA scheme obtains the least DoF performance. This phenomenon can be explained as that, on the one hand, the RIR scheme can aligns both ICI and IUI of which the RIA scheme only can align the later one, so that the RIR scheme gets higher DoF gain than RIA scheme. On the other hand, the B-DRIA scheme aligns both IUI and ICI, and meanwhile takes less slots to achieve IA, hence it keeps the highest DoF gain. In addition, we can note that, as the number of user antennas reaches 25, compared with the TDMA scheme, the DoF gain of the RIA scheme, RIR scheme and B-DRIA scheme are 11.54\%, 24.8\% and 56\%, respectively. Furthermore, it is worth noting that some points of the DoF keep constant locally when $N$ increases. This comes from the fact that, when the base station sends the same amount of information $\left\lceil {{M \mathord{\left/{\vphantom {M K}} \right.\kern-\nulldelimiterspace} K}} \right\rceil$ to each user, if $M$ cannot be divided to $K$, the base station does not send redundant symbols $[{M \mathord{\left/ {\vphantom {M K}} \right. \kern-\nulldelimiterspace} K} - \left\lceil {{M \mathord{\left/ {\vphantom {M K}} \right. \kern-\nulldelimiterspace} K}} \right\rceil $ separately. This measurement leads to no increase of the DoF and we call this operation as round down operation loss.

\begin{figure}[!t]
\centering
\includegraphics[width=2.5in]{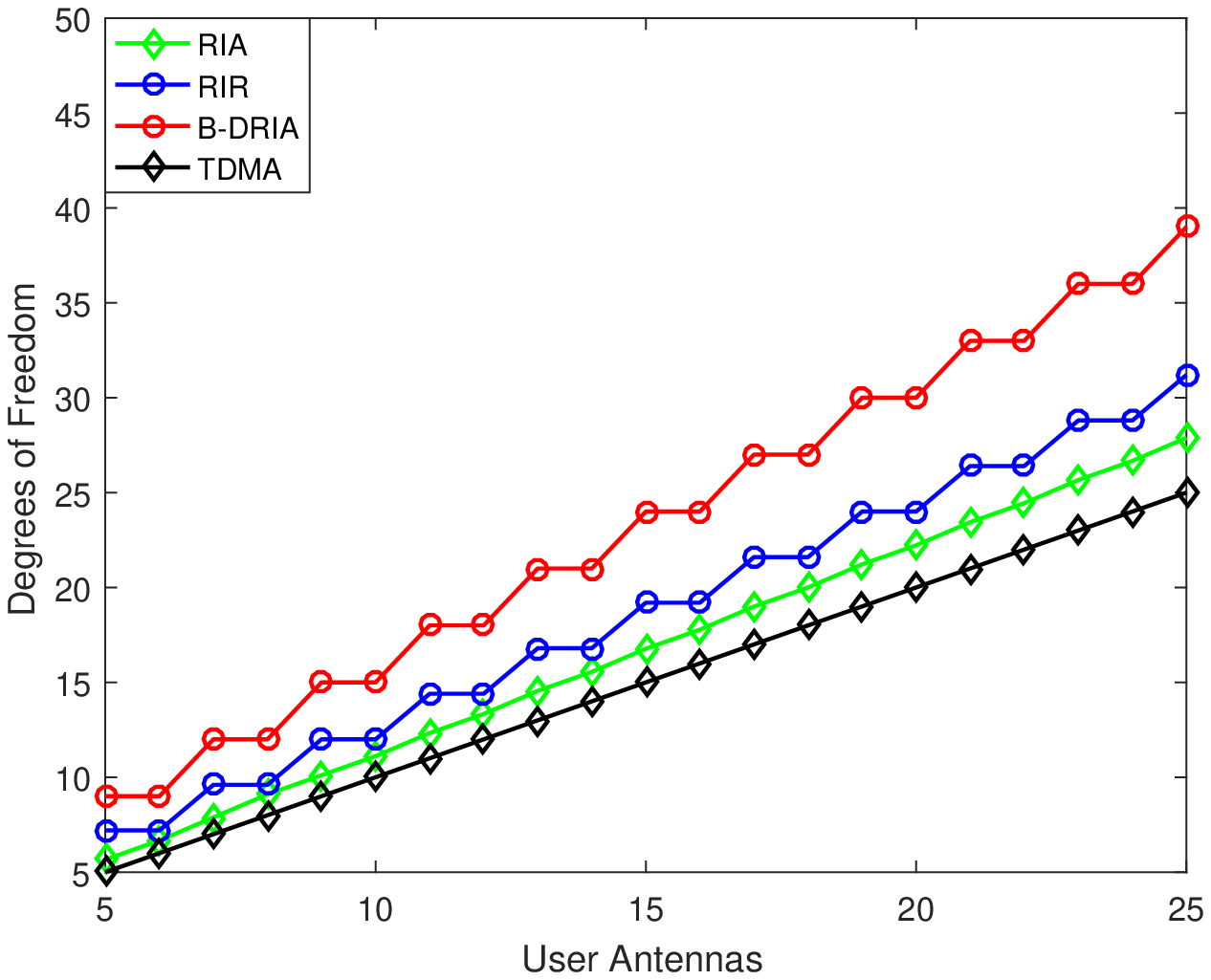}
\caption{DoF versus $N$ with fixed $\rho  = {3 \mathord{\left/{\vphantom {3 2}} \right.\kern-\nulldelimiterspace} 2}$ of the schemes.}
\label{Fig10}
\end{figure}

Then we further analyze how does the DoF is affected by the parameter $M$ under two special values of $\rho$, i.e., $\rho  = 1$ and $\rho  = 2$. For the former, we set the value of $\rho$ approaching to the critical point 2 and the other parameters used herein are the same as that used for the Fig. 10. Under this configuration, for the RIR scheme, $\varphi  = 1$ slots per group are used for signal transmission, and for the B-DRIA scheme, $\bar \varphi  = 0$ slot is used for signal transmission, in other words, B-DRIA scheme does not implement the second phase. It can be witnessed from the Fig. 11 that, the proposed two schemes achieve 36\% DoF gain than the benchmark schemes and meanwhile, both RIR scheme and B-DRIA scheme get the same DoF. The former confirms that the proposed schemes have a better effect on the improvement of DoF, and later is lies in the absence of the second phase so that the spatial gain is lost. It is also worth noting  that the RIA scheme has the same performance as the TDMA scheme. It can be explained as that RIA scheme transmits signals in the first slot and then transmits interference in the second slots where the whole process can be equivalent to the TDMA scheme with two slots. Notice that the third reason can also be used to explain why does the B-DRIA scheme not get the DoF gain from the past interference.
\begin{figure}[!t]
\centering
\includegraphics[width=2.5in]{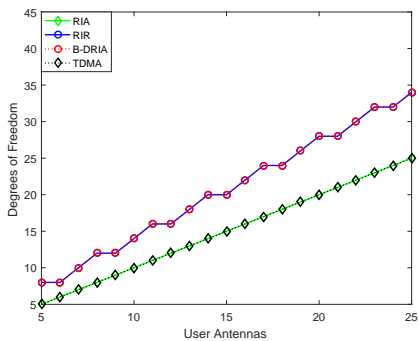}
\caption{DoF versus $N$ with fixed $\rho  = 2$ of the schemes.}
\label{Fig11}
\end{figure}

Finally, we set the value of $\rho$ approaching to the critical point 1 and the other parameters used herein are the same as that used for the Fig. 10. Under this configuration, for the RIR scheme and B-DRIA scheme, more than 2 slots per group and 1 slot per group are used for signal transmission, respectively. The Fig. 12 shows that the RIA scheme obtains 19.5\% DoF gain than the TDMA scheme, and meanwhile the performance of the two proposed schemes is polarized, i.e., the B-DRIA scheme acquires 100\% DoF gain than TDMA scheme while the performance of RIR scheme is degraded which is worse than RIA scheme. The former result can be explained as  that, for the B-DRIA scheme, with the multi-slots’ simultaneous transmission, the space resources have been fully utilized and the obtained DoF is tight to the upper bound \cite{5074376}. The later result is mainly because that, when   approaches 1, too many slots of each group are taken. In specific, when the number of user antennas reaches 25, there are $\bar \varphi {{ = }}26$ slots should be used for the single cell transmission which leads to serious waste of the time and space resources.

It is worth noting that, although the performance of the B-DRIA scheme is completely superior to the RIR scheme, in practice, the application scenario of the former is restricted. In specific, for the sake of DoA estimation, the location of the users should be stationary or moves within the current cell, otherwise the estimation should be reacquired which leads to the poor mobility of the network. Therefore, for the selection of the two proposed schemes, the choice should be made according to specific application scenario.

\begin{figure}[!t]
\centering
\includegraphics[width=2.5in]{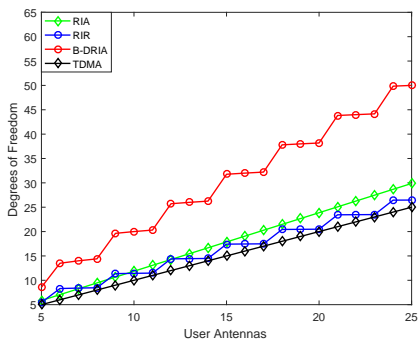}
\caption{DoF versus N with fixed $\rho  = 1$ of the schemes.}
\label{Fig12}
\end{figure}

\subsection{Performance Analysis with Fixed M}
We analyze how does the DoF is affected by the parameter $\rho$ under the condition that the parameter $M$ is fixed, i.e., $M = 72$. In specific, the configuration of the network is set to $L,M,K,N = (2,72,3,N)$, and meanwhile the value of the $N$ ranges from 36 to the 72. Aiming to this configuration the parameter goes through three intervals, i.e., ${3 \mathord{\left/{\vphantom {3 2}} \right. \kern-\nulldelimiterspace} 2} < \rho  \le 2$, ${4 \mathord{\left/
 {\vphantom {4 3}} \right. \kern-\nulldelimiterspace} 3} < \rho  \le {3 \mathord{\left/ {\vphantom {3 2}} \right. \kern-\nulldelimiterspace} 2}$ and $1 < \rho  \le {4 \mathord{\left/ {\vphantom {4 3}} \right. \kern-\nulldelimiterspace} 3}$ where the corresponding intervals measured by the parameter $N$ is $36 \le N < 48$ $48 \le N < 54$ and $54 \le N < 72$, respectively.

From the result shown in the Fig. 13, we find that, for the RIR scheme, the DoF keeps constant in each interval, while increasing sharply when the value of $N$ exceeds one critical point. This phenomenon is caused by round down operation loss mentioned before. Meanwhile, when the value of $N$ reaches 36, 48 and 54, the obtained DoF gains are 20\%, 14.3\% and 10.3\%, respectively. It indicates that, with the increase of $N$, the improvement made by RIR scheme becomes weaken. In another word, the DoF gain is inversely related to the critical points of $\rho$. Herein, the local optimum of DoF gain is obtained at critical points of $\rho$, and the global optimum of DoF gain is acquired when $\rho$ approaches 2.

\begin{figure}[!t]
\centering
\includegraphics[width=2.5in]{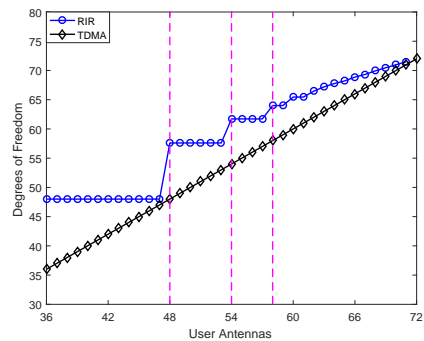}
\caption{DoF versus $N$ with fixed $M=72$ of RIR scheme.}
\label{Fig13}
\end{figure}

From the result shown in the Fig. 14, we discover that, for the B-DRIA scheme, the DoF keeps constant in each interval, while increasing sharply when the value of $N$ exceeds one critical point. The reason for this phenomenon is the same as that for the RIR scheme. Meanwhile, when the value of $N$ reaches 36, 48, and 54, the obtained DoF gains are 33.3\%, 50\% and 60\%, respectively. That is to say, the DoF gain is positively related to the critical points of $\rho$. Herein, the local optimum of DoF gain is obtained at the critical points of $\rho$, and the global optimum of DoF gain is acquired when $\rho$ approaches 1.

\begin{figure}[!t]
\centering
\includegraphics[width=2.5in]{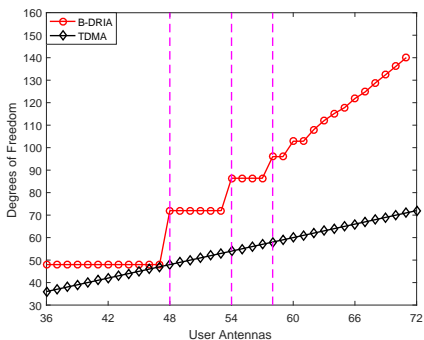}
\caption{DoF versus $\!N\!$ with fixed $M\!=\!72$ of B-DRIA scheme.}
\label{Fig14}
\end{figure}

\section{Conclusion}
In this paper, two new IA schemes were proposed for cellular $K$-user MIMO downlink networks, i.e., the RIR scheme and the B-DRIA scheme. For the former, the redundant symbols of the interference signals are eliminated with delayed CSIT and meanwhile, the desired symbols are provided to the target cell’s users without causing secondary interference. The analysis indicates that it achieves better DoF gain than the RIA scheme but, when the transceiver antennas ratio $\rho$ approaches 1, the performance improvement becomes weakened. Therefore, we further proposed the B-DRIA scheme, in which the beamforming technique is adopted to eliminate ICI and meanwhile, the distributed retrospective interference alignment is developed to use the past interference signals. Simulation results show that the B-DRIA scheme obtains larger DoF than the RIR scheme locally. Specifically, when $\rho$ approaches 1, two schemes obtain the same DoF. While as $\rho$ approaches 2, the DoF of the B-DRIA scheme is superior than the later.

\ifCLASSOPTIONcaptionsoff
  \newpage
\fi

\end{document}